\documentclass[12pt]{article}
\usepackage{amsmath, amssymb, amsthm, amsfonts}
\usepackage{paralist}
\newif\ifpdf
    \ifx\pdfoutput\undefined
    \pdffalse 
\else
    \pdfoutput=1 
    \pdftrue
\fi

\ifpdf
        \usepackage[pdftex]{graphicx}
        \DeclareGraphicsExtensions{.pdf, .jpg, .tif, .png}
\else
        \usepackage{graphicx}
        \DeclareGraphicsExtensions{.eps, .jpg}
\fi

\newcommand{\InsertFig}[4]
{\begin{figure}[ht]
      \centerline{
        \includegraphics[width=#4]{#1}
      }
      \caption{{\footnotesize  #2}}
      \label{#3}
\end{figure}
}

\newcommand{\Eq}[1]     {(\ref{#1})}

\newcommand{\Th}[1]     {Thm.~\ref{#1}}
\newcommand{\Lem}[1]    {Lem.~\ref{#1}}
\newcommand{\Cor}[1]    {Cor.~\ref{#1}}

\newcommand{\Fig}[1]    {Fig.~\ref{#1}}
\newcommand{\Sec}[1]    {\S\ref{#1}}
\newcommand{\Tbl}[1]    {Table~\ref{#1}}
\newcommand{\Prop}[1]   {Prop.~\ref{#1}}

\newcommand{\Tr}{\mathop{\rm Tr}}

\renewcommand{\mod}{\mbox{mod }}
\newcommand{\Fix}[1]{\mbox{Fix}(#1)}
\newcommand{\Sym}[1]{\mbox{Sym}(#1)}
\newcommand{\Rev}[1]{\mbox{Rev}(#1)}
\newcommand{\Grp}[1]{\langle #1 \rangle}
\newcommand{\R}{{\mathbb R}}
\newcommand{\Z}{{\mathbb Z}}

\newcommand{\C}{{\mathbb C}}
\newcommand{\Henon}{H\'enon }
\newcommand{\calA}{\mathcal{A}}
\newcommand{\calE}{\mathcal{E}}
\newcommand{\calG}{\mathcal{G}}
\newcommand{\calN}{\mathcal{N}}
\newcommand{\calO}{\mathcal{O}}
\newcommand{\calR}{\mathcal{R}}
\newcommand{\calS}{\mathcal{S}}
\newcommand{\calU}{\mathcal{U}}

\newcommand{\calRA}{\mathcal{R}_\mathcal{A}}
\newcommand{\calRE}{\mathcal{R}_\mathcal{E}}

\newcommand{\Raa} {$\bf{R_{AA}}$}
\newcommand{\Rae} {$\bf{R_{AE}}$}
\newcommand{\Ree} {$\bf{R_{EE}}$}
\newcommand{\Se}  {$\bf{S_E}$}
\newcommand{\Sa}  {$\bf{S_A}$}
\newtheorem{thm}{Theorem}
\newtheorem{lem}[thm]{Lemma}
\newtheorem{cor}[thm]{Corollary}
\newtheorem{prop}[thm]{Proposition}

\theoremstyle{definition}
\newtheorem{ex}{Example}[section]

\newtheorem*{defn}{Definition}

\newtheorem*{rem}{Remark}

\begin{document}

\title{Reversors and Symmetries for Polynomial Automorphisms of the Plane}
\author{A. G{\'o}mez\thanks{
       Support from the CCHE Excellence grant for Applied
       Mathematics  is gratefully acknowledged.}\\
       {\small \begin{tabular}{c}
       Department of Mathematics\\
       University of Colorado\\
       Boulder CO 80309-0395\\
       and\\
       Departamento de Matem\'aticas\\
       Universidad del Valle, Cali Colombia
       \end{tabular}}
       \and J. D. Meiss\thanks{ Support from NSF grant DMS-0202032 is
       gratefully acknowledged.  JDM would also like to thank H.
       Dullin and J. Roberts for helpful conversations.}\\
       {\small \begin{tabular}{c}
       Department of Applied Mathematics\\
       University of Colorado\\
       Boulder CO 80309-0526
       \end{tabular}}}\date{\today}
\maketitle

\begin{abstract}
     We obtain normal forms for symmetric and for reversible polynomial
automorphisms (polynomial maps that have polynomial inverses)
of the plane. Our normal forms are based on the generalized \Henon
normal form of Friedland and Milnor.  We restrict to the case that the
symmetries and reversors are also polynomial automorphisms.
We show that each such reversor has finite-order, and that
for nontrivial, real maps, the reversor has order 2 or 4.
The normal forms are shown to be unique up to finitely many choices.
We investigate some of the dynamical consequences of reversibility, especially for
the case that the reversor is not an involution.
\end{abstract}

\vspace*{1ex}
\noindent

\section{Introduction}\label{sec:intro}
Polynomial maps provide one of the simplest, nontrivial classes of
nonlinear dynamical systems. A subset of these, called {\em
polynomial automorphisms}, have polynomial inverses---thus these
maps are diffeomorphisms. Since this subset is closed under
composition, polynomial automorphisms form a group that we denote
by $\calG$. These maps can have quite complicated dynamics, as
exemplified by the renowned \Henon quadratic map
\cite{Henon69,Henon76}, which is in $\calG$.  The family of
generalized \Henon maps \cite{Friedland89},
\begin{equation}\label{eq:henon}
     h(x,y) = (y, p(y)-\delta x) \;,
\end{equation}
for any polynomial $p,$ is also in $\calG$ whenever $\delta \neq 0$, since
\[
    h^{-1}(x,y) = \left(\delta^{-1}(p(x)-y), x\right) \;,
\]
is also polynomial. These maps are area-preserving when $\delta = \pm 1$,
and orientation-preserving when $\delta > 0$.

The structure of $\calG$ is well understood, thanks to a classic result of
Jung \cite{Jung42}. In this paper we will make use of Jung's theorem to
investigate polynomial automorphisms that have symmetries or reversing
symmetries in $\calG$. Our results also extensively use the normal form for
polynomial automorphisms as compositions of generalized \Henon maps obtained
by Friedland and Milnor \cite{Friedland89}.

A diffeomorphism $g$ has a symmetry if it is conjugate to itself; that is, there exists
a diffeomorphism $S$ such that
\begin{equation}\label{eq:symmetric}
   g = S^{-1} \,g \,S \;.
\end{equation}
Similarly, $g$ has a reversing symmetry, or is ``reversible'', if
it is conjugate to its inverse \cite{Devaney76,Sevryuk86, Lamb94b, Lamb98b};
that is, there exists a diffeomorphism $R$ such that
\begin{equation}\label{eq:reversible}
       g^{-1} = R^{-1}\, g\,R \;.
\end{equation}
For example, generalized \Henon maps are reversible when $\delta = 1$, and they
have a nontrivial symmetry when there is an $\omega \ne 1$ such that
$p(\omega y) = \omega p(y)$. Reversible maps occur often in applications.
For example, reversibility often arises for Hamiltonian systems because
their phase spaces consist of coordinates $q$ and momenta $p$, and a
transformation that reverses the momenta, $R(q,p) = (q,-p)$, often corresponds
to reversal of time. Our goal in this paper is to classify the polynomial
automorphisms that satisfy \Eq{eq:symmetric} or \Eq{eq:reversible} with $S,R \in \calG$.

The basic properties of symmetries and reversors are discussed in \cite{Lamb92}
and reviewed in \cite{Lamb98b}. The set of symmetries of $g$ is never empty
since the identity always satisfies \Eq{eq:symmetric}. Moreover, since the
composition of any two symmetries is also a symmetry, they form a group
\begin{equation}\label{eq:symgroup}
     \Sym{g}=\{S \in \calG:S^{-1}\,g\,S=g\} \;.
\end{equation}
In group theory terminology, $\Sym{g}$ is the {\em centralizer} of $g$ in $\calG$.
Note that the family $\{g^{j}:\ j\in\Z\}$ is a subgroup of $\Sym{g}$. We will
say that $S$ is a nontrivial symmetry of $g$ in the case that $S \neq g^j$.
If all of the symmetries of $g$ are trivial then $\Sym{g}$ is isomorphic
to $\Z$ (or to $\Z_{n}$ if $g$ has finite order).

Similarly, whenever $R$ is a reversor for $g$, then so are each of the members of the family
\begin{equation}\label{eq:family}
     \{R_{j,k} = g^{j}R^{2k+1}:  j,k \in \Z \}\;,
\end{equation}
Thus if $R$ is a reversor, then its inverse is one as well. On the other
hand, the composition of any two reversors is a symmetry of $g$ (and is
not a reversor unless $g$ is an involution). Thus for example, if $R$
is a reversor then $R^2$ is a symmetry. Moreover, the composition of a
symmetry and of a reversing symmetry of $g$ yields a reversing symmetry.
It follows that the set of all symmetries and reversing symmetries of
$g$ is a group, usually referred as the group of {\em reversing symmetries} of $g$
\begin{equation}\label{eq:revgroup}
    \Rev{g}=\{ f \in \calG:f^{-1}\,g\,f = g^{\pm 1} \}.
\end{equation}
The group $\Sym{g}$ is a normal subgroup of $\Rev{g}$ and $\Rev{g}/\Sym{g}$
is isomorphic to $\Z_2$, the permutation group with two elements.
The properties of the group of reversing symmetries have been
investigated in several recent papers
\cite{Baake97,Goodson99,Baake01,Roberts03} as well as the papers
in the collection \cite{Lamb98a}.

Reversors arising in physical examples often are involutions, $R^{2} = id$, so that
$R$ generates a group $\Grp{R} = \{id,R\} \simeq \Z_2$. If $g$ possesses an involutory reversing
symmetry $R$, then $\Rev{g} = \Sym{g} \rtimes \Grp{R}$
\cite{Baake97};\footnote{
   Recall that the semidirect product $\calG = {\cal N} \rtimes {\cal M} $ where
   ${\cal N}$ is a normal subgroup of $\cal G$ is defined so that if $g = (n,m) \in \calG$
   then the product is $g_1\,g_2 = (n_1\,m_1\,n_2\,m_1^{-1}, m_1\,m_2)$.
   This contrasts with the
   direct product ${\cal N} \times {\cal M}$ where $g_1\,g_2 = (n_1\,n_2,m_1\,m_2)$.
}
therefore, in this case it suffices to determine the structure of $\Sym{g}$ to obtain a full
description of $\Rev{g}$.  However, reversors need not be involutions;
examples of maps with noninvolutory reversors (called weakly reversible
by Sevryuk \cite{Sevryuk86}) were given by Lamb \cite{Lamb92}.  As we
will see in \Sec{sec:reversors}, any reversor in $\Rev{g}$ has finite
order, $R^k = id$.  Note that if $k$ were odd, then $g$ would itself be
an involution and thus is dynamically trivial; so if we restrict to
nontrivial maps in $\calG$, then their reversors have even order.
Moreover, for real maps and reversors, we will show that the order must
be two or four.  After we submitted the current paper, we learned that
Roberts and Baake have also announced this result for real maps in
$\calG$ \cite{Roberts03}.

We will also demonstrate that if $g$ has a nontrivial symmetry, then
there is a subgroup of $\Sym{g}$ generated by an finite-order map and a
``root" of $g$.  In particular, we will show

\begin{thm}\label{thm:symmetryMain}
    Suppose $g$ is a polynomial automorphism of the plane that possesses
    nontrivial symmetries. Then $g$ is conjugate to a map of the form
    \[
        s (H) ^q \;,
    \]
    where $H$ is a composition of generalized \Henon maps \Eq{eq:henon}, $s$ is a diagonal
    linear map with finite order, $s^k = id$, and either $s \ne id$ or $q\ne 1$.
    The normal form has commuting symmetries $s$ and $H$,and
    $\Sym{G} \supset \Grp{H} \times \Grp{s} \simeq \Z \times \Z_k$.
\end{thm}
\noindent
This theorem is stated in a more detailed way in \Sec{sec:symmetries} as
\Th{thm:symmetryNormalForm} and \Cor{thm:symmetryCorollary}. Note that for the real case,
$s$ is at most order two (this result was also announced in \cite{Roberts03}).

In a previous paper we obtained normal forms for the automorphisms
that are reversible by an involution in $\calG$ \cite{Gomez02}.  Just
as in \cite{Friedland89}, these normal forms are constructed from
compositions of generalized \Henon maps; however, in this case, the
reversors are introduced by including two involutions in this
composition.  We showed there that the involutions can be normalized
to be either ``elementary'' involutions, or the simple affine
permutation
\begin{equation}\label{eq:t}
    t(x,y) = (y,x) \;.
\end{equation}
The second major result of the current paper is that in the general case,
reversible automorphisms also have at least two basic reversors that are also
either elementary or affine. In particular we will prove

\begin{thm}\label{thm:main}
    Suppose $g$ is a nontrivial reversible automorphism. Then $g$ possesses a reversor of
    order $2n$ in $\calG$ and is conjugate to one of the
    following classes:
    \begin{list}{}
        {\setlength{\labelwidth}{1cm}\setlength{\leftmargin}{2.5cm}
        \setlength{\labelsep}{1cm}
        \setlength{\topsep}{3ex}\setlength{\itemsep}{1.5ex}}
        \item[\Raa] ${\displaystyle\tau_{\omega}^{-1}\,H^{-1}\, \tau_{\omega}\,H,}$
        \item[\Rae]  $\tau_{\omega}^{-1}\,H^{-1}\, e_{2}\,H$,
        \item[\Ree]  $e_{1}^{-1}\,t\,H^{-1}\, e_{2}\,H\,t$,
    \end{list}
    where
    \begin{itemize}
        \item the map $H$ is a composition of generalized \Henon transformations \Eq{eq:henon},
        \item $\tau_{\omega}$ is the affine reversor, $\tau_\omega(x,y) = (\omega y,x)$, such that
              $\omega$ is a primitive $n^{\text{th}}$ root of unity, and
    \item the maps $e_{1},e_{2}$ are elementary reversors,
        \[
            e_i(x,y) =\left(p_i(y)-\delta_i\,x,\epsilon_i\,y\right)
        \]
            where
            $p_i(\epsilon_i\,y)=\delta_i\,p_i(y)$
            and $\epsilon_i^{2},\delta_i^{2}$ are primitive
            $n^{\text{th}}$ roots of unity.
    \end{itemize}
\end{thm}

\noindent
As we will see in \Sec{sec:reversors}, the reversor is an involution ($n=1$)
unless the polynomials in the \Henon maps and the elementary transformations satisfy
a common scaling condition. In this case we will see that $\omega$, $\epsilon_i^2$
and $\delta_i^2$ lie in subsets of the $n^{\text{th}}$ roots of unity
that we will explicitly construct.

We also show in \Sec{sec:reversors} that the maps in \Th{thm:main}
can be normalized, and that once this is done the normal forms are
unique up to finitely many choices. These details are contained in the more complete
statement, \Th{thm:normalforms}.

We finish the paper with a discussion of some examples and their dynamics.

\section{Background}
We start by giving some definitions and basic results concerning the algebraic
structure of the group $\calG$, presenting our notation, and reviewing the work
of Friedland and Milnor \cite{Friedland89}.

\subsection{Jung's Theorem}
The group $\calG$ is the group of {\em polynomial automorphisms} of the complex plane,
the set of bijective maps
\[
     g:(x,y)\rightarrow\left(X(x,y),Y(x,y)\right),\ \ X,Y \in \C[x,y] \;,
\]
having a polynomial inverse.  Here $\C[x,y]$ is the ring of
polynomials in the variables $x$ and $y$, with coefficients in $\C$.
In general, we consider this complex case, but in some instances we
will restrict to the case of real maps.  The {\em degree} of $g$ is
defined as the largest of the degrees of $X$ and $Y$.

The subgroup $\calE\subset \calG$ of {\em elementary} (or triangular)
maps consists of maps of the form
\begin{equation}\label{eq:elem}
     e:\left(x,y\right)\rightarrow \left(\alpha x+p(y), \beta y +\eta \right) \;,
\end{equation}
where $\alpha \beta \neq 0$ and $p(y)$ is any polynomial. The subgroup of
affine automorphisms is denoted by $\calA$. The affine-elementary maps will be
denoted by $\calS=\calA\cap \calE$.

We let $\hat{\calS}$ denote the group of diagonal
affine automorphisms,
\begin{equation}\label{eq:diagonal}
    \hat{s}:(x,y)\rightarrow \left(\alpha x +\xi, \beta y +\eta\right) \;.
\end{equation}
The group $\hat{\calS}$ is the largest subgroup of $\calS$
normalized by the permutation $t$ \Eq{eq:t}, i.e., such that
$t\,\hat{\calS}\,t=\hat{\calS}$. On the  other hand, the centralizer of $t$ in
$\calS$ is the subgroup of maps that commute with $t$
\[
     C_{\calS}(t)=\{s\in \calS:\,s\,t\,s^{-1}=t\} \;.
\]
These are the diagonal automorphisms \Eq{eq:diagonal} with $\alpha=\beta$
and $\xi=\eta$. Conjugacy by $t$ will be denoted by $\phi$,
\begin{equation}\label{eq:conjugacy}
    \phi(g)= t\,g\,t.
\end{equation}
Thus if $s\in C_{\calS}(t)$, then $\phi(s)=s$.

According to Jung's Theorem \cite{Jung42} every polynomial automorphism
$g\notin \calS$, can be written as
\begin{equation}\label{eq:Jung}
    g=g_m\, g_{m-1}\, \cdots \, g_2\, g_1,\quad\quad
    g_i\in \left(\calE\cup\calA\right)\setminus\calS,\ i=1,\ldots m \;,
\end{equation}
with consecutive terms belonging to different subgroups $\calA$ or $\calE$. An
expression of the form \Eq{eq:Jung} is called a {\em reduced word} of length
$m$. An important property of a map written in this form is that its degree
is the product of the degrees of the terms in the composition
\cite[Thm. 2.1]{Friedland89}. A consequence of this fact is that the
identity cannot be expressed as a reduced word
\cite[Cor. 2.1]{Friedland89}. This means that $\calG$ is the free
product of $\calE$ and $\calA$ amalgamated along $\calS$. The structure of $\calG$ as an
amalgamated free product determines the way in which reduced words
that correspond to the same polynomial automorphism, are related.

\begin{thm}\label{thm:uniqueness} (cf. \cite[Cor. 2.3]{Friedland89}, or
    \cite[Thm. 4.4]{Magnus66}) Two reduced words
    $g_{m}\,\cdots\,g_{1}$ and $\tilde{g}_{n}\,\cdots\,\tilde{g}_{1}$
    represent the same polynomial automorphism $g$ if and only if $n=m$ and there
    exist maps $s_{i}\in \calS,\ i=0,\dots ,m$ such that $s_{0}=s_{m}=id$ and
    $\tilde{g_{i}}=s_{i}\,g_{i}\,s_{i-1}^{-1}$.
\end{thm}

From this theorem it follows that the length of a reduced word \Eq{eq:Jung} as
well as the degrees of its terms are uniquely determined by $g$. The sequence
of degrees $(l_{1},\dots,l_{n})$ corresponding to the maps $(g_{1},\dots
,g_{m})$, after eliminating the $1$'s coming from affine terms, is referred to
as the {\em polydegree} of $g$.

A map is said to be {\em cyclically-reduced} in the trivial case
that it belongs to $\calA\cup\calE$ or when it can be written as a
reduced word \Eq{eq:Jung} with $m\ge2$ and $g_m$, $g_1$ not in the
same subgroup $\calE$ or $\calA$.

\subsection{Conjugacies of Polynomial Autormorphims}

Two maps $g,\tilde{g}\in \calG$ are conjugate in $\calG$ if there
exists $f\in\calG$ such that $g=f\,\tilde{g}f^{-1}$.  If $f$
belongs to some subgroup $\mathcal{F}$ of $\calG$ we say that $g$
and $\tilde{g}$ are $\mathcal{F}$-conjugate.  It can be easily
seen that every $g\in\calG$ is conjugate to a cyclically-reduced
map.  Moreover, an explicit calculation shows that every affine
map $a$ can be written as $a=s\,t\,\tilde{s}$, where $t$ is given
by \Eq{eq:t} and $s,\tilde{s}$ are affine-elementary maps.  From these
facts it follows that every polynomial automorphism is either {\em
trivial} (i.e., conjugate to an elementary or an affine map) or is
conjugate to a reduced word of the form,
\begin{equation}\label{eq:te-form}
    g=t\,e_m\,\cdots\,t\,e_{2}\,t\,e_1,\quad e_i\in \calE\setminus\calS,\quad
    i=1,\cdots m,\quad m\geq 1.
\end{equation}
Moreover this representative of the conjugacy class is
unique up to modifications of the maps $e_i$ by diagonal affine automorphisms
and cyclic reordering. More precisely we have the following theorem
(following \cite[Thm. 4.6]{Magnus66}).

\begin{thm}\label{thm:conjugate}
    Two nontrivial, cyclically-reduced words $g=g_{m}\,\dots\,g_{1}$ and
    $\tilde{g}=\tilde{g}_{n}\,\dots\,\tilde{g}_{1}$ are $\calG$-conjugate if and only if
    $m=n$ and there exist automorphisms
    $s_{i}\in \calS,\ i=0,\dots,m$ with $s_{m} \equiv s_{0}$, and a cyclic permutation,
    \[
        \left(\hat{g}_{m},\ldots ,\hat{g}_{1}\right)=
        \left(\tilde{g}_{k},\ldots,\tilde{g}_{1},
        \tilde{g}_{m},\ldots,\tilde{g}_{k+1}\right)
    \]
    such that $\hat{g}_{i}=s_{i}\,{g}_{i}\,s_{i-1}^{-1}$. In that case,
    \[
    s_{0}\,g\,s_{0}^{-1}=\hat{g}_{m}\,\ldots\,\hat{g}_{1}.
    \]

    In particular, if $g=t\,e_m\,\cdots\, t\,e_1$ and
    $\tilde{g} = t\,\tilde{e}_{m}\,\cdots\, t\,\tilde{e}_{1}$ are conjugate, there
    exist diagonal automorphisms $s_{i}\in\hat{\calS},\ s_{m} \equiv s_{0}$, and a cyclic
    reordering,
    \[
    \left(\hat{e}_{m},\ldots ,\hat{e}_{1}\right)=
     \left(\tilde{e}_{k},\ldots ,\tilde{e}_{1},
     \tilde{e}_{m}\ldots,\tilde{e}_{k+1}\right),
    \]
    such that $t\,\hat{e}_{i}=s_{i}\,t\,e_{i}\,s_{i-1}^{-1}$ and
    \[
    s_{0}\,g\,s_{0}^{-1}=t\,\hat{e}_{m}\,\cdots\,t\,\hat{e}_{1}.
    \]
\end{thm}
\begin{proof}
Let $g=g_{m}\,\dots\,g_{1}$ and
$\tilde{g}=\tilde{g}_{n}\,\dots\,\tilde{g}_{1}$ be two nontrivial,
cyclically-reduced, conjugate words. By assumption, there is a reduced word
$f=f_{k}\,\cdots\,f_{1}\in \calG$, such that $g=f\,\tilde{g}\,f^{-1}$. Then,
\begin{equation}\label{eq:conjugate}
    g_{m}\,\cdots\,g_{1}=
    f_{k}\,\cdots\,f_{1}\,\tilde{g}_{n}\,\cdots\,\tilde{g}_{1}\,
    f_{1}^{-1}\,\cdots\,f_{k}^{-1}.
\end{equation}
However, the word on the right hand side of \Eq{eq:conjugate} is not reduced.
Since $\tilde{g}$ is cyclically-reduced, we can suppose, with no loss of
generality, that $f_{1}$ and $\tilde{g}_{1}$ belong to the same subgroup $\calA$
or $\calE$, so that $f_{1}$ and $\tilde{g}_{n}$ lie in different subgroups.
Taking into account \Th{thm:uniqueness} and that \Eq{eq:conjugate} represents a
cyclically-reduced map, we can reduce to obtain
\begin{equation}\label{eq:reduced}
    \tilde{g}_{n}\,\cdots\,\tilde{g}_{1}\,f_{1}^{-1}\,\cdots\,f_{k}^{-1}\,=
    \begin{cases}
    \tilde{g}_{n}\,\cdots\,\tilde{g}_{k+1}\,\tilde{s}_{k} &\text{if }n\geq k\\
    \tilde{s}_{n}\,f_{n+1}^{-1}\,\cdots\,f_{k}^{-1} &\text{if } n< k,
    \end{cases}
\end{equation}
where $\tilde{s}_{n}, \tilde{s}_{k}\in \calS$. Moreover there exist
$\tilde{s}_{i}\in \calS,\ \tilde{s}_{0}=id$,
such that
$\tilde{g}_{i}\,\tilde{s}_{i-1}\,f_{i}^{-1}=\tilde{s}_{i},$
for $i=1,\dots ,\min(n,k)$.

For the case $n\geq k$,
\begin{align*}
    g_{m}\,\dots\,g_{1}&=
    f_{k}\,\cdots\,f_{1}\,\tilde{g}_{n}\,\cdots\,\tilde{g}_{k+1}\,\tilde{s}_{k}\\
    &=(\tilde{s}_{k}^{-1}\,\tilde{g}_{k})\,\tilde{g}_{k-1}\,\cdots\,\tilde{g}_{1}\,
    \tilde{g}_{n}\,\dots\,\tilde{g}_{k+2}\, (\tilde{g}_{k+1}\,\tilde{s}_{k}),
\end{align*}
and applying \Th{thm:uniqueness} we have the result. The case $n<k$ follows
analogously.

To prove the second statement of this theorem it is enough to recall that
given $s\in \calS$, $t\,s\,t$ stays in $\calS$ if and only if $s$ is diagonal.
\end{proof}

This implies that the length of a
cyclically-reduced word is an invariant of the conjugacy class. Since a
nontrivial, cyclically-reduced word has the same number of elementary and
affine terms, we refer to this number as the {\em semilength} of the word.
\Th{thm:conjugate} also implies that two cyclically-reduced maps that are
conjugate have the same polydegree up to cyclic permutations. We will call
this sequence the polydegree of the conjugacy class.

\subsection{Generalized \Henon transformations}
A {\em generalized \Henon transformation} is
any map of the form \Eq{eq:henon} where $\delta\neq 0$ and $p(y)$ is a polynomial
of degree $ l \ge 2$. Notice that a generalized \Henon transformation can be written
as the composition
\[
   h = t\,e  \;, \quad e(x,y) = (p(y)-\delta x, y) \;.
\]
If $p(y)$ has leading coefficient equal to $1$
($\pm 1 $ in the case of real automorphisms)
and center of mass at $0$,
\begin{equation}\label{eq:normalPoly}
     p(y)=y^{l}+O(y^{l-2}) \;,
\end{equation}
we say that the polynomial is {\em normal}, and consequently that the \Henon
transformation is {\em normalized}. In \cite{Friedland89},
Friedland and Milnor obtained normal forms for conjugacy classes of elements in
$\calG$, in terms of generalized \Henon transformations.
\begin{thm}\label{thm:friedland} \cite[Thm. 2.6]{Friedland89}
    Every nontrivial $g \in \calG$ is conjugate to a composition of
    generalized \Henon transformations, $h_{m}\,\cdots\,h_{1}$. Additionally it
    can be required that each of the terms $h_{i}$ be normalized and in that case
    the resulting normal form is unique, up to finitely many choices.
\end{thm}
To prove this result it is enough to take \Th{thm:conjugate} into account and
check that, given a map $g=t\,e_{m}\,\cdots\,t\,e_{i}$, it is possible to
choose diagonal affine automorphisms $s_{i},\ i=1,\dots ,m$, in such a
way that $s_{m}$ coincides with $s_{0}$, and for every $i$,
$s_{i}\,t\,e_{i}\,s_{i-1}^{-1}$ is a normalized \Henon transformation.
In the next section we will use the following generalization of \Th{thm:friedland}.
\begin{lem}\label{lem:henon}
    Given a cyclically-reduced map of the form \Eq{eq:te-form}, there exist
    diagonal affine automorphisms $s_{m},s_{0}\in C_{\calS}(t)$, such that
    \[
    s_{m}\,g\,s_{0}^{-1}=h_{m}\,\cdots\,h_{1},
    \]
    where every term $h_{i}$ is a normal \Henon transformation.
    The \Henon maps are unique up to finitely many choices.
\end{lem}
\begin{proof}
Consider a cyclically reduced map \Eq{eq:te-form}, with
\[
   t\,e_{i}: (x,y)\rightarrow\left(\beta_{i}\,y+\eta_{i},\alpha_{i}\,x+p_{i}(y)\right),
      \qquad i=1,\dots ,m.
\]
We look first for diagonal affine maps $s_{i},\ i=0,\dots ,m$, with
$s_{0},s_{m}\in C_{\calS}(t)$ and such that the maps
$t\,\hat{e}_{i}=s_{i}\,t\,e_{i}\,s_{i-1}^{-1}$ are H\'enon transformations. If
we denote
\[
     s_{i}(x,y)=(u,v)=(a_{i}x+b_{i},c_{i}y+d_{i}),
\]
the problem reduces to the set of equations,
\begin{equation*}
    a_{0}=c_{0},\qquad b_{0}=d_{0},\qquad a_{m}=c_{m},\qquad b_m=d_{m},
\end{equation*}
and
\begin{equation*}
    \beta_{i} a_{i}=c_{i-1},\qquad b_{i}=d_{i-1}-\eta_{i} a_{i},
    \qquad i=1,\dots, m.
\end{equation*}
This system can be easily solved in terms of $2m$ parameters; a particular
solution is obtained by choosing
$c_{0}=\dots=c_{m-1}=1$ and $d_{0}=\dots=d_{m-1}=0$.
We can assume now that the terms $t\,e_{i}$ in \Eq{eq:te-form} are already
\Henon maps, but not necessarily in normal form. In that case,
\begin{equation*}
    t\,e_{i}:\left(x_{i-1},x_{i}\right)\rightarrow
    \left(x_{i},x_{i+1}\right),\quad x_{i+1}=p_i(x_i)-\delta_{i}x_{i-1}.
\end{equation*}
Setting $s_{i}\left(x_{i},x_{i+1}\right)=(y_{i},y_{i+1}),\ y_i=a_i x_i +b_i$
and $t\,\hat{e}_{i}=s_{i}\,(t\,e_{i})\,s_{i-1}^{-1}$, we have,
\begin{gather*}
    t\,\hat{e}_{i}:\left(y_{i-1},y_{i}\right)\rightarrow
    \left(y_{i},y_{i+1}\right),\quad
    y_{i+1}=\hat{p}_{i}(y_{i})-\hat{\delta}_{i}y_{i-1},\\
    \hat{p}_{i}(y)=a_{i+1}p_i\left(\frac{y-b_{i}}{a_{i}}\right)+\text{const.},\quad
    \hat{\delta}_{i}=\frac{a_{i+1}}{a_{i-1}}\delta_{i}.
\end{gather*}
In order to have leading coefficients equal to $1$ we need
\[
     \kappa_{i} a_{i+1}=a_{i}^{l_{i}}, \quad i=1,\dots ,m,
\]
where $\kappa_{i}$ is the leading coefficient of $p_{i}$. On the other hand we
require $a_{m+1}=a_{m},\ a_{1}=a_{0}$, since by assumption $s_{m},s_{0}\in
C_{\calS}(t)$. It is easy to see that these conditions yield $a_{1}$ up to
$l^{th}$-roots of unity, where $l=l_{1}\cdots l_{m-1}(l_{m}-1)$. All other
$a_{i}$ are then uniquely determined. Finally, the coefficients
$b_{i},\ i=1,\dots ,m$, can be chosen so that the next to highest order terms
are equal to zero and we set $b_{0}=b_{1}$ and
$b_{m+1}=b_{m}$ to ensure that $s_{0},s_{m}$ are in the centralizer of $t$.

The above arguments also show that the terms $t\,\hat{e}_{i}$ are
unique up to replacing the polynomials $\hat{p}_{i}(y)$ and
parameters
$\hat{\delta}_{i}$ with
$\zeta^{l_{i}\dots l_{0}}\hat{p}_{i}(y/\!\zeta^{l_{i-1}\dots l_{0}})$ and
$\zeta^{l_{i}l_{i-1}}\hat{\delta_{i}}$ respectively, where $\zeta$ is any
$l\text{th-}$ root of unity and $l_{0}=1$.
\end{proof}

\subsection{Roots of Unity}
As the proof of \Lem{lem:henon} shows, the normal forms for polynomial
automorphisms are unique only up to a scaling by certain roots
of unity. In subsequent sections, we will see that symmetric and reversible
automorphisms are associated with several subgroups of the roots of
unity. In anticipation of these results, we provide some notation for these subgroups.

Let $\calU$ be the group of all roots of unity in $\C$ (the points with
rational angles on the unit circle) and
$\calU_{n}$ be the group of $n^{th}$-roots of unity:
\begin{equation}
   \calU_{n} \equiv \{ z\in\C : z^n = 1\} \;.
\end{equation}

Given a sequence of normal \Eq{eq:normalPoly} (nonlinear) polynomials, $p_{1}(y),\dots ,p_{m}(y)$,
and some root of unity $\zeta$, define the set
$\calR(\zeta) =\calR(\zeta;p_{1}(y),\dots,p_{m}(y))$, by
\begin{equation}\label{eq:Rzeta}
    \calR(\zeta) \equiv
   \{\omega\in \C: \omega\,p_{2i+1}(\omega\,y)=\zeta\,p_{2i+1}(y),\
    \omega\,p_{2i}(\omega\,y)=p_{2i}(\zeta\,y),\ 1\leq 2i,2i+1\leq m\} \;.
\end{equation}
It can be observed that if $\omega\in \calR(\zeta)$ then
$\omega^{-1}\in\calR(\zeta^{-1})$, while if
$\omega_{1}\in\calR(\zeta_{1})$ and $\omega_{2}\in\calR(\zeta_{2})$ it follows
that $\omega_{1}\,\omega_{2}\in\calR(\zeta_{1}\,\zeta_{2})$. Thus, the set
\begin{equation}\label{eq:RE}
    \calRE \equiv \bigcup_{\zeta\in \calU}\calR(\zeta) \;,
\end{equation}
is a subgroup of $\calU$. Moreover, unless the sequence of polynomials reduces
to one monomial, there are only finitely many $\zeta\in\calU$ such that
$\calR(\zeta)$ is nonempty. In this case $\calRE$ has finite order, hence it
coincides with one of the groups $\calU_{n}$. On the other hand
given any $\omega\in\calRE$,  the definition \Eq{eq:Rzeta} implies that there is a unique
\begin{equation}\label{eq:zetaOmega}
  \zeta(\omega) =  \zeta  \mbox{ such that } \omega \in \calR(\zeta)\; .
\end{equation}

Now we define the set
\begin{equation}\label{eq:RA}
\calRA \equiv \{\omega\in \calU :p_{i}(\omega\,y)=\omega\,p_{i}(y),\ i=1,\dots ,m\} \;.
\end{equation}
It can be seen that $\calRA$ is a subgroup of the group of roots of unity
$\calU$.
Moreover, $\omega\in\calRA$ {\em iff} $\zeta(\omega)=\omega^{2}$, i.e.\ {\em iff}
$\omega\in\calR(\omega^{2})$, so that $\calRA$ is a subgroup of
$\calRE$. Note that $\calRA$ has finite order for any nonempty sequence of
(nonlinear) polynomials.

Finally we define the subgroup $\calN$ of $\calRE$ by
\begin{equation}\label{eq:N}
    \calN \equiv
        \left\{\omega\in\C:\omega\,p_{i}(\omega \, y)=\zeta\,p_{i}(y),
        \text{ for some }\zeta\in\calRA\right\}=
        \bigcup_{\zeta\in\calRA}\calR(\zeta) \;,
\end{equation}
and the subgroup of $\calRA$ defined by
\begin{equation}
    \calN' \equiv \{\zeta\in \calRA: \calR(\zeta) \neq \emptyset \} \;.
\end{equation}

These groups have the ordering
\begin{equation}\label{eq:Np}
         \calN' \subseteq \calRA \subseteq \calN \subseteq \calRE \subseteq \calU \;.
\end{equation}

\begin{ex}\label{ex:poly}
Given the three normal polynomials
\[
     p_1(y) = y^5 \;,\quad p_2(y) = y^{13}+a\,y^5 \;,\quad p_3(y) = y^{21} \;,
\]
then if $a \neq 0$ we find that $\calR(\zeta)$ is nonempty only for $\zeta \in \calU_4$ and that
\[
       \calN' = \calRA = \calU_4 \subset \calN = \calRE = \calU_8 \;.
\]
On the other hand, if $a=0$ then
\[
       \calN' = \calRA = \calU_4 \subset \calN = \calU_8 \subset
       \calRE=\calU_{16}\;.
\]
\end{ex}

\section{Symmetric Automorphisms}\label{sec:symmetries}

In this section we investigate the structure of cyclically reduced
words that represent polynomial automorphisms possessing
nontrivial symmetries. We will show that if $g$ is a nontrivial
polynomial automorphism with nontrivial symmetries, then $g$ is
conjugate to a map of the form $s\,H^q$. In this form,
$s$ is a finite-order, affine-elementary symmetry and $H$
is a cyclically reduced symmetry. This decomposition of $g$ gives
rise to a subgroup of $\Sym{g}$ isomorphic to $\Z \times \Z_{n}$
where $n$ is the order of $s$. Similar subgroups have been found
for the particular case of polynomial mappings of generalized
standard form \cite{Roberts03}.

Using \Th{thm:friedland} we can, with no loss of generality, assume that $g$ is
in \Henon normal form.
\begin{thm}\label{thm:symmetryNormalForm}
Suppose that $g=h_m \, h_{m-1}\,\cdots\, h_1$ is a polynomial
automorphism  in \Henon normal form with a  nontrivial symmetry in
$\calG$. Then there exist diagonal linear transformations $s,
\tilde{s}$ such that $s$ has finite order and
\begin{equation}\label{eq:symmetricNormalForm}
      g         = s^{j}\,(\tilde{g})^{q} \;, \mbox{ where }
      \tilde{g} = \tilde{s}\,h_{r}\,\cdots\,h_{1} \;;
\end{equation}
where $s,\tilde{g}$ are commuting symmetries of $g$, $m=qr$, and
either $s\ne id$ or $q\ne 1$.
\end{thm}
\begin{proof}
By assumption there is a nontrivial symmetry, $f$, and by the
condition \Eq{eq:symmetric}, since $g$ is cyclically reduced, $f$
must be either an affine-elementary map or a nontrivial cyclically
reduced word. The set $\{ g^{j}\,f^{k}: j,k \in \Z \}$ is a
subgroup of $\Sym{g}$. By replacing $f$ with some convenient
element in that subgroup if necessary and using
\Th{thm:conjugate}, we may assume that
$f=s_m\,h_{k}\,\cdots\,h_{1},$  with $s_m$ a diagonal affine map,
and $0\leq k<m$. In this case the relation $f\,g\,f^{-1}=g$
becomes
\begin{equation*}
    s_m\,h_{k}\,\cdots\,h_{1}\,h_{m}\,\cdots\,h_{k+1}\,s_m^{-1}=
    h_{m}\,\cdots\,h_{1}.
\end{equation*}
Since $h_i = t\,e_i$, \Th{thm:uniqueness} then implies the
existence of diagonal affine maps $s_{i}, \; i = 0 \ldots m-1$,
with $s_0 = s_m$ such that
\begin{equation}\label{eq:modk}
    h_{i+k}=s_{i}^{-1}\,h_{i}\,s_{i-1},
\end{equation}
where the indices should be understood $\mod m$.

Let $r=\gcd(m,k)$ the greatest common divisor of $m$ and $k$, and
define integers $q,p$ such that $m=qr,$ $k=pr$. In this case there
exist integers $j$ and $a$ such that $r=jk - am$ where we may
assume $j,a>0$ \cite{Jacobson85}. It then follows that $r\equiv
jk\ \mod m,$ so that $h_{r+i}=h_{jk+i}$. Iterating \Eq{eq:modk} we
obtain
\[
     h_{i+ r}=s_{i+(j-1)k}^{-1}\,\cdots\,s_{i+k}^{-1} s_{i}^{-1}\,
     (h_{i})\,s_{i-1}s_{i-1+k}\,\cdots\,s_{i-1 + (j-1)k} \;.
\]
Defining $\tilde{g}_n = h_{n r} h_{nr-1}\,\cdots\,h_{(n-1)r+1}$, we then obtain
\begin{equation}\label{eq:r-piece}
  \tilde{g}_{n+1}= \tilde{s}_n^{-1} \, \tilde{g}_n\, \tilde{s}_{n-1}
   \;, \mbox{ for } n=1,\ldots,q-1 \;,
\end{equation}
where $\tilde{s}_n = s_{nr}s_{nr+k}\,\cdots\, s_{nr + (j-1)k}$.
Since $m = qr$, we can use induction on \Eq{eq:r-piece} to obtain
\begin{equation}\label{eq:g-symmetric}
     g = \tilde{s}_{q-1}^{-1}\,\cdots \tilde{s}_1^{-1}\tilde{s}_0^{-1}\,\tilde{g}^{q}
\end{equation}
where
\begin{equation}
  \tilde{g}=  \tilde{s}_0 \,h_{r}\,\cdots\,h_{1}=
  \tilde{s}_{0}\,\tilde{g}_{1} \;.
\end{equation}
The leading affine-elementary map in \Eq{eq:g-symmetric} is
actually the $j$-th power of a simpler map since $r \equiv jk \
\mod m$ implies that $\tilde{s}_n = s_{njk}s_{(nj+1)k}\,\cdots\,
s_{((n+1)j - 1)k}$. Using this we regroup the $q$ groups of $j$
terms as $j$ groups of $q$ to obtain
\begin{align*}
 \tilde{s}_0\tilde{s}_1\,\cdots\,\tilde{s}_{q-1}
   &= (s_{0}s_{k}\,\cdots\,s_{(j-1)k})(s_{jk}\,\cdots\, s_{(2j-1)k})
          \quad\cdots\quad(s_{(q-1)jk}\,\cdots s_{(jq-1)k})&\\
   &= (s_{0}s_{k}\;\;\cdots\;s_{(q-1)k})(s_{qk}\;\;\cdots\; s_{(2q-1)k})
            \,\cdots\,(s_{(j-1)qk}\;\;\cdots\;s_{(jq-1)k})&\\
   &= (s_{0}s_{k}\;\;\cdots\,s_{(q-1)k})^{j}
\end{align*}
since $qk \equiv 0 \ \mod m$.
Thus we have shown that whenever $g$ has a symmetry it has the form
\Eq{eq:symmetricNormalForm} with
\[
     s \equiv s_{(q-1)k}^{-1}\,\cdots\,s_{k}^{-1}\,s_{0}^{-1}.
\]

We still have to prove that $s\in\Sym{g}$. Using \Eq{eq:modk} and the fact that
$h_{m+i}=h_{i}$ allows us to obtain
\[
    h_{i}=s_{m+i-k}^{-1}\,\cdots\,s_{m+i-qk}^{-1}\,(h_{m+i-qk})\,
             s_{m+i-1-qk}\,\cdots\,s_{m+i-1-k}.
\]
As $qk\equiv 0\ \mod m$ the above relations imply
\[
     s_{m+r-qk}\,\cdots\,s_{m+r-k}\,(h_{r}\,\cdots\,h_{1})
     = (h_{r} \,\cdots\,h_{1})\,s_{m-qk}\,\cdots\,s_{m-k},
\]
which readily yields $s\,\tilde{g}=\tilde{g}\,s$.
To see that $s$ must be of finite order, note that since $g$ is assumed to be
in normal \Henon form, the relation
\[
s\,h_{m}\,\cdots\,h_{1}\,s^{-1}=h_{m}\,\cdots\,h_{1}
\]
means that $s$ is a diagonal linear transformation
\[
s:(x,y)\rightarrow (a_{0}x,a_{1}y),
\]
that conjugates a normal form to itself. Then, as in
\Lem{lem:henon}, the scaling factors $a_{0},a_{1}$ must be roots
of unity, so that $s$ has finite order. Similarly since each of
the $s_i$ are diagonal, so is $\tilde{s}_0$.

To finish we show that if $f$ is a nontrivial symmetry then either
$s\ne id$ or $q\ne 1$. To see this we observe that \Eq{eq:r-piece}
gives
\begin{align*}
     f &= s_{m}\,h_{k}\,\cdots\,h_{1} \;,\\
       &= s_{m}(\,\tilde{s}_{p-1}^{-1}\,\cdots\, \tilde{s}_0^{-1})\,
           (\tilde{s}_{0}\,h_{r}\,\cdots\,h_{1})^{p} \;,\\
       &= s^{a}\,\tilde{g}^{p} \;,
\end{align*}
where we used  $k=pr$ and that $jk\equiv r\ \mod m$. Therefore if
$q= 1$ then $m = r = k$ so that $j = p = 1$, so that when $s=id$
then $f=\tilde{g} = g$ is a trivial symmetry.
\end{proof}

Note that since $s$ has finite order and is a  diagonal linear map, then if it
is real it must be an involution.

From this theorem we see that every polynomial automorphism that
possesses nontrivial symmetries and is not the $q$-fold iteration
of a nontrivial automorphism must have nontrivial symmetries
conjugate to affine-elementary maps. In the case of a map given in
\Henon normal form it is not difficult to establish conditions
under which the group of such symmetries is nontrivial.

\begin{prop}\label{thm:affineSym}
    If $s$ is an affine-elementary symmetry of the \Henon normal form
    map $g=h_{m}\,\cdots\,h_{1}$, then $s = s_\omega$ where
    \begin{equation}\label{eq:normalsymm}
         s_\omega(x,y)  = \left(\frac{\zeta(\omega)}{\omega} x,\omega y\right), \quad \omega \in \calU_k
    \end{equation}
    where $\calU_k =  \calRE$ if $m$ is even, and $\calU_k =  \calRA$ if $m$ is odd. Thus the
    set of affine elementary symmetries is a cyclic group isomorphic to $\calU_k$.
  \end{prop}
\begin{proof}
According to \Th{thm:uniqueness}, $s\in\calS$ is a symmetry if and
only if there exist maps $s_{i}\in\hat{\calS},$ $i=0,\ldots,m,$
$s_{0}=s_{m}=s,$ such that
\begin{equation}\label{eq:symcond}
    s_{i}\,h_{i}\,s_{i-1}^{-1}=h_{i} \;.
\end{equation}
Moreover, given that $g$ is in normal form, each of the maps $s_{i}$ must be
of the form $s_{i}(x,y) =  (a_{i}x,a_{i+1}y)$.
In that case \Eq{eq:symcond} translates into the conditions
\begin{gather*}
    a_{i-1}=a_{i+1},\quad a_{0}=a_{m} \;,\\
    p_{i}(a_{i}x)=a_{i+1}p_{i}(x) \;.
\end{gather*}
When $m$ is even, we must have $a_{2k+1} = \omega$ and $a_{2k} = \zeta(\omega)/\omega$
for $\omega\in \calRE$ and $\zeta(\omega)$ is defined by \Eq{eq:zetaOmega}.
Thus $s = s_0$ has the promised form. Since $\zeta(\omega^2) = (\zeta(\omega))^2$,
the set of symmetries $s\in\calS$ is a cyclic group isomorphic to $\calRE$.

In a similar way if $m$ is odd, all $a_{i}$ must be equal to some
$\omega\in\calRA$, \Eq{eq:RA}, so that $s(x,y) = (\omega x,\omega y)$.
Moreover since $\zeta(\omega) = \omega^2$ for $\omega \in \calRA$, the
symmetries $s\in\calS$ are of the promised form, and they form a cyclic
group isomorphic to $\calRA$.
\end{proof}

Finally, as a corollary to these results we can prove a more complete statement of \Th{thm:symmetryMain}.

\begin{cor}\label{thm:symmetryCorollary}
    Suppose $g$ is a polynomial automorphism of the plane that possesses
    nontrivial symmetries. Then $g$ is conjugate to a map of the form
    \[
        s_\omega \,(H) ^q \;,
    \]
    where $H = h_n\,h_{n-1}\ldots h_1$ is a composition of normal
    generalized \Henon maps \Eq{eq:henon} and $s_{\omega}$ is
    given by \Eq{eq:normalsymm}
    where  $\omega \in \calRE$ if $nq$ is even and $\omega \in
    \calRA$ if $nq$ is odd. The normal form has commuting
    symmetries $s_\omega$ and $H$, and either $\omega \ne 1$ or $q \ne 1$.
\end{cor}
\begin{proof} According to \Th{thm:symmetryNormalForm}, $g$ is conjugate to $s^j(\tilde{s}H)^q$.
Using \Th{lem:henon}, we can conjugate this map with a new diagonal-affine transformation to
normalize the map $\tilde{s}H$. The conjugacy commutes with $s^j$ since both are diagonal,
and according to \Prop{thm:affineSym}, $s^j$ has the promised form.
\end{proof}
\section{Reversible automorphisms}\label{sec:reversors}
In \cite{Gomez02} we described conjugacy classes for polynomial
automorphisms that are reversible by involutions.  Although the
involution condition appears in a natural way in many cases and it
was originally one of the ingredients in the definition of
reversible systems \cite{Devaney76}, this requirement can be
relaxed.  Indeed, many of the features still hold in the more
general case (see, e.g., \cite{Lamb98b} for further discussion).
Moreover, all maps with an involutory reversor are reversible in
the more general sense. Finally, it is certainly true that there
exist reversible maps that do not possess any involution as a
reversor; for example \cite{Lamb92}. We begin by showing that the
general reversor in $\calG$ is conjugate to one that is affine or
elementary. Then we will show that every reversor in $\calG$ is of
finite order. Finally we prove the main theorem.

\subsection{Affine and Elementary Reversors}

\begin{lem}\label{lem:elem}
    If $g \in \calG$ is nontrivial and reversible, then it is conjugate to an automorphism
    $\tilde g$ that has an elementary or an affine reversor. If the semilength
    of $g$ is odd, then $\tilde g$ has an affine reversor.
\end{lem}
\proof
Following \Eq{eq:te-form}, we may assume that
$g=t\,e_{m}\,\cdots\,t\,e_{1}$ is written as a cyclically reduced
word.  Similarly let $R$ be the reduced word for a reversing symmetry
of $g$.  Since both $g$ and $g^{-1} = R^{-1}\,g\,R$ are cyclically
reduced, then unless $R$ is in $\calA \cup \calE$, it cannot be a
nontrivial cyclically reduced map. Recall that if $R$ is a reversor for $g$ so
are the maps $R\,g^{j}$ and that they have the same order as $R$ if
this order is even.  Thus without loss of generality, we can assume
that $R$ is shorter than $g$; otherwise as in the proof of
\Th{thm:conjugate}, there is an $\tilde R= R \,g^{j}$ for some $j\in
\Z$, such that $\tilde{R}$ shorter than $g$ and is also a reversor.
Moreover, by replacing $R$ with $R\,g$ if necessary, we can assume
that
\begin{equation}\label{eq:min}
    R=s_{0}^{-1}\,e_{k}\,t\,\cdots\,t\,e_{1},\ s_{0}\in\hat{\calS} \;,
    \quad k<m,\ s_{0}\in\hat{\calS} \;.
\end{equation}
Following \Th{thm:conjugate} and given that $g=R\,g^{-1}\,R^{-1}$, there
must exist maps $s_{i}\in\hat{\calS}$ such that
$t\,e_{i^{*}}^{-1}=s_{i}\,t\,e_{i}\,s_{i-1}^{-1}$ for $i^{*}=k-i+1$
and $i=1,\dots ,m$, where the indices are taken modulus $m$.  Consider
then the map $\hat{g} = f\,g\,f^{-1}$, where $f =
t\,e_{\nu}\,\cdots\,t\,e_{1}$, and $k=2\nu$ or $k=2\nu + 1$.  Then a
simple calculation shows that $\hat g$ has a reversor $\hat R =
f\, R\,f^{-1}$ that is either $s_\nu^{-1}t$ when $k$ is
even or $s_\nu^{-1}e_{\nu+1}$ when $k$ is odd.  Thus $\hat g$ is
conjugate to $g$ and has a reversor in $\calA \cup \calE$.

Now suppose that the semilength of the conjugacy class is odd and that
$g=t\,e_{2k+1}\,\cdots\,t\,e_{1}$ has an elementary reversor,
$R=s_{0}^{-1}\,e_{1}$. Reordering terms we see that the map
\begin{equation*}
    \hat{g}=
    t\,e_{k+1}\,\cdots\,t\,e_{2}\,t\,e_{1}\,t\,e_{2k+1}\,\cdots\,t\,e_{k+2}
\end{equation*}
has an affine reversing symmetry $\hat{R}$, conjugate to some reversor in the
family $R\,g^{j}$. Therefore in the case of conjugacy classes of odd
semilength we may always assume affine reversors.
\qed

\begin{rem}
Along the same lines of the proof of the previous lemma it can be seen
that if $R$ is any reversor of a cyclically reduced, nontrivial
automorphism $g$, then $R$ can be written as $R_{0}\,g^{j}$ for some
$j\in \Z$ and $R_{0}$ shorter than $g$. Since
$(R_{0}\,g^{j})^{2\mu+1}\,g^{l}=R_{0}^{2\mu+1}\,g^{j+l}$ it follows that
the reversors generated by $R_{0}\,g^{j}$ form a subset of the family
generated by $R_{0}$.

We can go further by conjugating with maps in $S$ and replacing $g$ by
its inverse if necessary, so that $g$ can be taken to have the form \Eq{eq:te-form} and
$R_{0}$, the form \Eq{eq:min}. Then, a short calculation shows
that the reversors generated by $R_{0}$ are of the form
\[
R=R_{0}^{2\mu+1}\,g^{j}=(s_{0}^{-1}\,t\,s_{k}^{-2\mu}\,t\,s_{0}^{-1})\,
        e_{k}\,t\,\cdots\,t\,e_{1}\,g^{j}, \quad s_{0},s_{k}\in\hat{\calS}
\]
whenever the index $\mu\ge 0$. For negative $\mu$ a similar
result follows; however if the reversor has finite order, it
is enough to consider only one of these possibilities. Now, since
$s_{0},s_{k}\in\hat{\calS}$ the term inside
the parenthesis in the above expression reduces to an elementary affine
map. When $j\ge 0$ no further simplifications are possible, so that the
reduced word that represents $R$ has length $2k-1,$ if the length is considered
modulus $2m$. For $j<0$ an additional simplification yields
\[
    R=(s_{0}^{-1}\,t\,s_{k}^{-2\mu}\,t\,s_{0}^{-1})\,
        t\,e_{k+1}^{-1}\,\cdots\,e_{m}^{-1}\,t\,g^{j+1} \;,
\]
so that, modulus $2m$,  the length of the word becomes $2(m-k)+1$.
We can conclude that if any two reversors of a nontrivial,
cyclically reduced map belong to a common family, then their
modulus $2m$ lengths must coincide or be complementary (i.e.\
their sum is equal to $2m$).
\end{rem}

\subsection{Finite-Order Reversors}

We now show that reversors in $\calG$ have finite order. Recall that when $R$ is a
reversing symmetry for $g$, each of the maps \Eq{eq:family}
is also a reversor and that whenever that $R$ has finite order, the order must
be even, unless $g$ is an involution.

\begin{thm}\label{thm:finite}
    Every polynomial reversor of a nontrivial, polynomial automorphism has
    finite even order. In the case of real transformations the order is $2$ or $4$.
\end{thm}
\begin{proof}
Following \Th{thm:friedland}, we may assume that $g$ is in \Henon normal form
and from \Lem{lem:elem} that $R$ is an affine or an
elementary reversing symmetry for $g$. Now, it is easy to see that if $R$ is
affine, $R=s_{0}^{-1}\,t$, and if $R$ is elementary, $R=s_{0}^{-1}\,e_{1}$,
for some $s_{0}\in \hat{\calS}$, and $e_1(x,y) = (p_1(y)-\delta_1 x,y)$ a
normalized elementary map. The condition $g=R\,g^{-1}\,R^{-1}$ is then
equivalent to the existence of diagonal linear maps,
$s_{i}(x,y)=(a_{i}\,x,a_{i+1}\,y)$,\ $s_{m}=s_{0}$, such that
\begin{equation}\label{eq:transform}
  t\,e_{i^{*}}^{-1}=s_{i}\,t\,e_{i}\,s_{i-1}^{-1} \;,
    \quad \mbox{where} \; \left\{
        \begin{array}{ll}
          i^{*} = m-i+1, &  R \mbox{ affine} \\
          i^{*} = m-i+2, & R \mbox{ elementary}  \\
        \end{array} \right. \;.
\end{equation}
Here the indices are understood modulus $m$. This in turn means that
\begin{equation}\label{eq:deltas}
\begin{array}{l}
  \delta_{i}\,\delta_{i^{*}} =
         \displaystyle{\frac{a_{i-1}}{a_{i+1}}=\frac{a_{(i+1)^{*}}}{a_{(i-1)^{*}}}}\\
  p_{i^{*}}(a_{i}\,y)  =\delta_{i^{*}}\,a_{i+1}\,p_{i}\left(y\right) \\
  \end{array}
    \;, \quad i = 1,\ldots, m \;.
\end{equation}
Defining $\omega_{i}=a_{i}\,a_{i^{*}}$, then \Eq{eq:deltas} implies
\begin{equation}\label{eq:omegas}
\begin{array}{l}
    \omega_{i}=\omega_{i^{*}} \;,\quad
    \omega_{i-1}=\omega_{i+1}\\
    p_{i}(\omega_{i}\,y)=\omega_{i-1}\,p_{i}\left(y\right) \\
\end{array} \;,\quad i=1,\dots ,m \;.
\end{equation}
Thus all the odd $\omega_i$ are equal, as are all the even $\omega_i$.
Furthermore $\omega_{m} \equiv\omega_{0}$. It follows that when $R$ is affine
or when the semilength of $g$ is odd, all of the $\omega_{i}$ coincide.  Then
\Eq{eq:omegas} implies that all $\omega_{i} $ are primitive roots of unity of
the same order. The proof is complete upon noting that
$R^{2}(x,y)=(x/\!\omega_{0},y/\!\omega_{1})$ which means that $R$ has order
$2n$, where $n$ is the order of $\omega_{i}$. It can be observed that \Eq{eq:omegas}
implies that $2n$ must be a
factor of $2\,(l_{i}\,l_{i-1}-1)$, for all indices $i$. Finally we see that $R$
is real only if  $\omega_{0},\omega_{1}=\pm 1$, so that the order of $R$ is $2$
or $4$.
\end{proof}

It is not hard to find normal forms for elementary and affine reversors.

\begin{defn}[Normalized Elementary Reversor] An elementary map of the form
\begin{equation}\label{eq:e-reversor}
    e : (x,y)\rightarrow\left(p(y)-\delta\,x,\epsilon\,y\right),
    \qquad p(\epsilon\,y)=\delta\,p(y),
\end{equation}
with $p(y)$ a normal polynomial and $\epsilon^{2},\delta^{2}$ some primitive
$n^{th}$-roots of unity, will be called a {\em normalized elementary
reversor} of order $2n$. Note that $e^2 = (\delta^2 x, \epsilon^2 y)$.
\end{defn}

\begin{defn}[Normalized Affine Reversor]
Given any primitive $n^{th}$-root of unity $\omega$, the map
\begin{equation}\label{eq:a-reversor}
    \tau_{\omega}:(x,y)\rightarrow\left(\omega\,y,x\right)
\end{equation}
is a {\em normalized affine reversor} of order $2n$. Note
that $\tau_\omega^2 = (\omega x,\omega y)$
\end{defn}

\noindent
These normal forms will form the building blocks for the conjugacy classes of reversible
automorphisms.

Let us suppose now that the map $g=h_{m}\,\cdots\,h_{1}$ is in \Henon normal
form and that it has either an elementary or an affine reversing symmetry of order $2n$. In that case  the proof of \Th{thm:finite} implies there exist some primitive $n^{th}$-roots of unity, $\omega_{0}$ and $\omega_{1}$, that solve \Eq{eq:omegas}. Comparing this with \Eq{eq:Rzeta}, we see that $\omega_{1}\in\calR(\zeta)$, for $\zeta=\omega_{0}\,\omega_{1}$. Since $\omega_{1}$ generates $\calU_{n}$, and $\calRE$, \Eq{eq:RE}, is a group containing $\calR(\zeta)$, this implies that $\calU_n \subseteq \calRE$. If in addition the reversor is affine, then $\omega_{0}=\omega_{1}$ so that $\omega_{1}\in\calRA$, \Eq{eq:RA}, which implies that $\calU_{n} \subseteq \calRA$.

Reversibility imposes stronger conditions than are apparent in \Eq{eq:omegas} for some cases.
This occurs for words with even semilength that have an elementary reversor, and for words
with odd semilength (in this case, as we noted in \Lem{lem:elem} we can assume
that there is an affine reversor). We note that $i^{*}=i$ for
$i=k+1$ if the reversor is affine and $m=2k+1$, while this identity follows
for $i=1,k+1$ if the reversor is elementary and $m=2k$. For such indices,
\Eq{eq:deltas} implies the existence of some constants
$\hat{\epsilon}_{i}$ and $\hat{\delta_{i}}$ such that
\begin{equation}\label{eq:epsilons}
    p_{i}(\hat{\epsilon}_{i}\,y)=\hat{\delta}_{i}\,p_{i}(y),\quad
    \hat{\epsilon}_{i}^2=\omega_{i},\quad \hat{\delta}_{i}^{2}=\omega_{i-1},
\end{equation}
where the constants $\omega_{i}$ also satisfy \Eq{eq:omegas} for the
corresponding indices.

We can use \Eq{eq:omegas} and \Eq{eq:epsilons} to construct reversible
maps.  Let us suppose, for example, that we have $k$ normalized \Henon
transformations, $h_{1}(y),\dots ,h_{k}(y)$, a normal polynomial
$p_{k+1}(y)$, and a $n^{th}$-root of unity, $\omega\in
\calRA(p_{1}(y),\dots ,p_{k+1}(y))$, such that \Eq{eq:epsilons} holds
for $i=k+1$, if we set $\omega_{k}=\omega_{k+1}=\omega$.  Then it is
possible to choose the coefficient $\delta_{k+1}$ and the remaining
\Henon transformations, in such a way that the map
$g=h_{2k+1}\,\cdots\,h_{1}$ has an affine reversing symmetry of order
$2n$.  Furthermore the number of possible choices is finite.  As
similar statements follow in the other cases, we have a way to
generate all conjugacy classes for reversible automorphisms.  These
conditions also enables us to give an explicit description of
conjugacy classes for reversible automorphisms, as well as to provide
normal forms, as we show next.

\subsection{Normal Form Theorem}

We are now ready to prove the main result, which was given in the introduction as \Th{thm:main}.
Given the previous lemmas, we can now restate the result here in more detail:
\begin{thm}\label{thm:normalforms}
Let be $g$ a nontrivial automorphism that possesses a reversor of order $2n$,
and $\omega$ any primitive $n^{th}$-root of unity.
Then $g$ is conjugate to a cyclically reduced map of one of the following
classes:

\begin{list}{}
    {\setlength{\labelwidth}{1cm}\setlength{\leftmargin}{2.5cm}
    \setlength{\labelsep}{1cm}
    \setlength{\topsep}{3ex}\setlength{\itemsep}{1.5ex}}
    \item[\Raa]
        ${\displaystyle\tau_{\omega}^{-1}\,(h_{1}^{-1}\,\cdots\,h_{k}^{-1})\,
        \tau_{\omega}\,(h_{k}\,\cdots\,h_{1})} \;,\quad
        \omega \in \calRA(p_{1}(y),\dots ,p_{k}(y))$
    \item[\Rae]
        $\tau_{\omega}^{-1}\,(h_{1}^{-1}\,\cdots\,h_{k}^{-1})\,
            e_{k+1}\,(h_{k}\,\cdots\,h_{1}) \;,\quad
        \omega \in \calRA(p_{1}(y),\dots ,p_{k+1}(y))$,
    \item[\Ree]
        $e_{1}^{-1}\,(t\,h_{2}^{-1}\,\cdots\,h_{k}^{-1})\,
            e_{k+1}\,(h_{k}\,\cdots\,h_{2}\,t) \;,\quad
        \omega \in \calRE(p_{1}(y),\dots ,p_{k+1}(y))$,
\end{list}
where
\begin{itemize}
    \item the maps $h_{i}$ are normalized \Henon transformations,
    \item $\tau_{\omega}$ is the normalized affine reversor \Eq{eq:a-reversor},
    and
    \item if $(\omega_{i})$ is the sequence defined by
        $\omega_{1} = \omega,\ \omega_{2} = \zeta(\omega)/\!\omega,\
        \omega_{i+1} = \omega_{i-1}$, then the maps $e_{1},e_{k+1}$ are normal
        elementary reversors \Eq{eq:e-reversor}, with
        $\epsilon_{i}^{2} = \omega_{i},\ \delta_{i}^{2} = \omega_{i-1}$.

\end{itemize}
Furthermore these normal forms are unique up to finitely many choices.

Conversely, $\tau_{\omega}$ is a reversing symmetry for any map having normal
form \Raa\ or \Rae\, while $e_{1}$ is an elementary reversor for any map of the
form \Ree.
\end{thm}
\begin{proof}
We consider the conjugacy class of a polynomial automorphism
$g=t\,e_{m}\,\cdots\,t\,e_{1}$, having a reversing symmetry $R_{0}$, of order
$2n$. We may assume that $g$ is given in \Henon normal form and according to
\Lem{lem:elem} that $R_{0}$ is affine or elementary.
Moreover, when $m$ is odd it may be assumed that $R_{0}$
is affine. Then $R_{0}$ is of the form $s_{0}^{-1}\,t$ or $s_{0}^{-1}\,e_{1}$,
for some scaling $s_{0}:(x,y)\rightarrow(a_{0}\,x,a_{1}\,y)$. Throughout the
present discussion we continue using the notation introduced in
\Th{thm:finite}. In particular we know that the polynomials $p_{i}(y)$ satisfy
\Eq{eq:omegas}. Moreover if $R_0$ is affine and $m=2k+1$ the polynomial
$p_{k+1}(y)$ also satisfies \Eq{eq:epsilons}, while if $R_0$ is elementary and
$m=2k$, this condition is satisfied by $p_{1}(y)$ and $p_{k+1}(y)$.

We discuss the case $R_{0}$ affine; the case that
$R_{0}$ is elementary follows in a similar way. Note that if $R_{0}$ is
affine then, $R_{0}=\hat{s}_{0}\,\tau_{\omega}^{-1}\,\hat{s}_{0}^{-1}$,
for $\omega=a_{0}\,a_{1}$ and some diagonal linear map $\hat{s}_{0}$.
Letting $m=2k$ or $m=2k+1$, then \Eq{eq:transform} implies that
\begin{equation*}
    g=(s_{0}^{-1}\,t)\,\left(e_{1}^{-1}\,t\,\cdots\,t\,e_{k}^{-1}\,
        [t\,e_{k+1}^{-1}]\,s_{k}\,t\,e_{k}\,\cdots\,t\,e_{1}\right) =
        R_{0}\,R_{1},
\end{equation*}
where the term in brackets is absent if $m=2k$. Notice that $R_{1}$ is also a
reversing symmetry of order $2n$, conjugate to either $s_{k}\,t$ if $m$ is even,
or to $e_{k+1}^{-1}\,s_{k}=\phi(s_{k+1})\,e_{k+1}$ when $m$ is odd. In the
first of these cases we also note that there exists a diagonal linear map
$\hat{s}_{k}$ such that
$s_{k}\,t=\hat{s}_{k}\,\phi(\tau_{\omega})\,\hat{s}_{k}^{-1}$.
In the second case, note that the map $\tilde{e}_{k+1}=e_{k+1}^{-1}\,s_{k}$
is an elementary reversor of the form \Eq{eq:e-reversor}, except that the
polynomial $p_{k+1}(y)$ may not be normalized.
Therefore $g$ is $\calS$-conjugate to a map of the form
\begin{equation}\label{eq:normalform}
    \tau_{\omega}^{-1}\,e_{1}^{-1}\,t\,\cdots\,t\,e_{k}^{-1}\,t\,[e_{k+1}]
    \,[\tau_{\omega}]\,t\,e_{k}\,\cdots\,t\,e_{1},
\end{equation}
where some of the maps $t\,e_{i}$ have been modified, but only by scalings of
their variables, so that the polynomials $p_{i}(y)$ still have center of mass at
$0$. Furthermore $e_{k+1}$ is a (not necessarily normalized) elementary reversor
of the form \Eq{eq:e-reversor}, with
$\epsilon_{k+1}^{2}=\delta_{k+1}^{2}=\omega$. Finally, the brackets indicate
the terms that may be omitted, depending on $m$ odd or even.

We can now replace each of the maps $t\,e_{i},\ i=1,\dots,k$ with normalized
\Henon transformations, as well as the maps $e_{i}^{-1}\,t$ with the
corresponding inverses, by applying \Lem{lem:henon} to
$f=t\,e_{k}\,\cdots\,t\,e_{1}$. In this case the conjugating maps turn out to
be linear transformations. Note that the diagonal linear maps that commute with
$t$ also commute with any $\tau_{\omega}$ and that the only effect of the
conjugacies we apply on $e_{k+1}$ is to rescale the polynomial $p_{k+1}(y)$. In
this way we obtain a map of the form \Eq{eq:normalform}, conjugate to $g$,
where each of the terms $t\,e_{i},\ i=1,\dots,k$, is a normalized \Henon
transformation. For the even semilength case this already shows that $g$ is
conjugate to a map of the form \Raa, for some $n^{th}$-root of unity
$\omega\in \calRA(p_{1}(y),\dots ,p_{k}(y))$.

When $m$ is odd, we still have to normalize $e_{k+1}$ to
make the leading coefficient of $p_{k+1}(y)$ equal to $1$. This can be
achieved by choosing some convenient scalings
\begin{equation}\label{eq:scaling}
    s_{i}:(x,y)\rightarrow (a_{i}\,x,a_{i+1}\,y),\quad i=0,\dots k,
\end{equation}
$\phi(s_{0})=s_{0}$, to replace $t\,e_{k+1}$ with
$\phi(s_{k})\,t\,e_{k+1}\,s_{k}^{-1}$, and then each of the terms
$t\,e_{i}$ with $s_{i}\,t\,e_{i}\,s_{i-1}^{-1}$, while the
corresponding $t\,e_{i}^{-1}$ are replaced by
$\phi(s_{i-1})\,t\,e_{i}^{-1}\,\phi(s_{i}^{-1}) =
\phi(s_{i}\,t\,e_{i}\,s_{i-1}^{-1})^{-1}$.  It is not hard to see that
appropriate coefficients $a_{i}$ can be chosen to give the normal
form, and that they are unique up to $l\text{th-}$roots of unity, with
$l=l_{1}\,\cdots\,l_{k-1}\,(l_{k}\,l_{k+1}-1)$.

We still need to show that we can replace $\omega$ in these expressions with any
given root of unity of order $n$, and that the forms thus obtained are
uniquely determined up to finitely many possibilities.

Let us consider the case of normal form \Raa.
Using \Th{thm:conjugate} we see that the terms $t\,e_{i}$ can be modified only
by scalings of the variables, since we require those terms to stay normal.
If we apply to the terms $t\,e_{i}^{-1}$ the images
under the isomorphism $\phi$ of the transformations we use to modify
$t\,e_{i}$, as we did to obtain the normal forms, the structure of the
word is preserved and the parameter $\omega$ does not change either.
In the more general case we may replace $t\,e_{i}$ with
$t\,\hat{e}_{i}=s_{i}\,t\,e_{i}\,s_{i-1}^{-1}$,\ $s_{i}$ given by
\Eq{eq:scaling}, while for $i=2,\dots,k-1$,\ $t\,e_{i}^{-1}$ is replaced by
$t\,\hat{e}_{i}^{-1}=\tilde{s}_{i-1}\,t\,e_{i}^{-1}\,\tilde{s}_{i}^{-1}$,
\begin{equation*}
    \tilde{s}_{i}:(x,y)\rightarrow(\tilde{a}_{i+1}\,x,\tilde{a}_{i}\,y),\quad
    i=1,\dots,k-1.
\end{equation*}
In this case, and if $k\ge2$, it follows that
$t\,(e_{k}^{-1}\,t\,\tau_{\omega})$ must be replaced by
$t\,(\hat{e}_{k}^{-1}\,t\,\tau_{\hat{\omega}})=
\tilde{s}_{k-1}\,t\,e_{k}^{-1}\,t\,\tau_{\omega}\,s_{k}^{-1}$
while $\tau_{\omega}^{-1}\,e_{1}^{-1}$ becomes replaced with
$\tau_{\hat{\omega}}^{-1}\,\hat{e}_{1}^{-1}=
s_{0}\,\tau_{\omega}^{-1}\,e_{1}^{-1}\tilde{s}_{1}^{-1}$. If $k=1$\ we have to
replace $\tau_{\omega}^{-1}\,e_{1}^{-1}\,t\,\tau_{\omega}$ with
$\tau_{\hat{\omega}}^{-1}\,e_{1}^{-1}\,t\,\tau_{\hat{\omega}}=
s_{0}\,\tau_{\omega}^{-1}\,e_{1}^{-1}\,t\,\tau_{\omega}\,s_{1}^{-1}$.

For the structure of the word to remain unchanged we need
\begin{equation}\label{eq:lambdas}
    \tilde{a}_{1}=a_{0},\quad
    \lambda_{i-1}=\lambda_{i+1},\quad \text{and}\quad
    p_{i}(\lambda_{i}\,y)=\lambda_{i+1}\,p_{i}(y),
\end{equation}
where $i$ runs from $1$ to $k$,\ $\tilde{a}_{k+1}$ is defined to be equal to
$a_{k}$, and $\lambda_{i}=\tilde{a}_{i}/a_{i}$ for $i=1,\dots,k+1$, while
$\lambda_{0}$ is just defined to be equal to $\lambda_{2}$. We also note that
then $\hat{\omega}=\lambda_{1}\,\lambda_{2}\,\omega$. Now, for the map
$\hat{g}$ obtained in this way, to be in normal form, it is also necessary that
$\hat{\omega}$ lies in $\calRA=\calRA(\hat{p}_{1}(y),\dots,\hat{p}_{k}(y))=
\calRA(p_{1}(y),\dots,p_{k}(y))$.
It follows that, if we set
$\zeta=\lambda_{1}\lambda_{2}=\hat{\omega}/\!\omega$,\ $\zeta$ is also an
element of the group $\calRA$.

We thus have that the solutions $(\lambda_{1},\lambda_{2})$ of
\Eq{eq:lambdas}, yielding alternative normal forms for $g$, are of the
form $(\lambda,\zeta/\!\lambda)$, for some $\zeta\in\calRA$ and
$\lambda\in\calR(\zeta)$.
Therefore to obtain all possible normal forms it suffices to consider all
$\lambda\in\calN$ \Eq{eq:N}, and set $\lambda_{1}=\lambda,\
\lambda_{2}=\zeta(\lambda)/\!\lambda$. The requirement that the polynomials
$\hat{p}_{i}(y)$ have leading coefficients equal to $1$ allows to determine the
coefficients $a_{i},\tilde{a}_{i}$ up to $l\text{th-}$roots of unity, for
$l=l_{1}\,\cdots\,l_{k-1}\,(l_{k}-1)$. If $\calN=\calU_{d}$ \Eq{eq:Np}, all possible
normal forms arise by taking $a_{1}$ as any $(ld)\text{th-}$root of unity, and
$\lambda$ as $a_{1}^{l}$.

We show now that $\hat{\omega}$ can be chosen as any primitive
$n^{th}$-root of unity. Note that the possible $\hat{\omega}$ are of the
form $\zeta\,\omega$ for some $\zeta\in\calN'$. We know that for any
$\omega\in\calRA$, $\calR(\omega^{2})$ is a nonempty set, since it contains
$\omega$. Let us denote by $\calRA^{2}$ the subgroup of $\calRA$ that consists
of elements of the form $\omega^{2},\ \omega\in\calRA$.
It then follows that $\calRA^{2}$ is a subgroup of $\mathcal{N'}$. Now, given
that $\calRA=\calU_{r}$ for some $r$, it is not difficult to see that
$\calRA^{2}$ is a maximal subgroup of $\calRA$ if $r$ is even while
$\calRA=\calRA^{2}$, if $r$ is odd. In the last case, and in general whenever
$\mathcal{N'}=\calRA$, we see that $g$ can be written in normal form with
$\omega$ replaced by any $r\text{th-}$root of unity, and in particular by any
primitive $n^{th}$-root of unity.

However if $\calRA$ has even order, it is possible that $\mathcal{N'}$ reduces
to $\calRA^{2}\ne\calRA$ and then it is no longer clear that $\hat{\omega}$
can be chosen as an arbitrary $n^{th}$-root of unity. A short calculation
shows that $\hat{\omega}$ can still be taken as any $n^{th}$-root
of unity as long as the number $r/\!n$ be even. When $r/\!n$ is odd the only
admissible $n^{th}$-roots of unity are the numbers $\exp\frac{i2\pi\nu}{n}$,
with $\nu$ odd. In particular all primitive $n^{th}$-roots of unity are
still possible, but $\hat{\omega}$ cannot be taken equal to $1$, i.e.\ $g$
lacks involutory reversing symmetries associated to this normal form. It is still
possible that there exist involutory reversors, however corresponding to a
different reordering of the terms, in the case that the map has other
families of reversing symmetries.

In the case of normal form \Rae\ condition \Eq{eq:lambdas} should hold for
$i=1,\dots,k+1$ (although $\tilde{a}_{k+1}$ is not necessarily equal to
$a_{k}$), if $\lambda_{k+2}$ is defined to be equal to $\lambda_{k}$. We still
have $\hat{\omega}=\lambda_{1}\,\lambda_{2}\,\omega$, but we need in addition
that
\begin{equation*}
    \hat{\epsilon}_{k+1}=\lambda_{k+1}\,\epsilon_{k+1},\quad
    \hat{\delta}_{k+1}=\lambda_{k}\,\delta_{k+1},\quad
    \hat{\epsilon}_{k+1}^{2}=\hat{\delta}_{k+1}^{2}=\hat{\omega}.
\end{equation*}
These conditions imply that the solutions $(\lambda_{1},\lambda_{2})$ for
\Eq{eq:lambdas}, yielding normal forms for $g$ must be of the form
$\lambda_{1}=\lambda_{2}=\lambda$ for some $\lambda\in\calRA$. In other words
the set $\calN$ coincides with $\calRA$, while
the corresponding $\mathcal{N'}$ equals $\calRA^{2}$.
The statements about admissible $\hat{\omega}$ then follows, basically
unchanged.

We can make analogous considerations for normal form \Ree. The possible
normal forms are obtained in this case by considering any $\lambda\in\calRE$.
If we set $\lambda_{1}=\lambda$ and $\lambda_{2}=\zeta(\lambda)/\!\lambda$,
the remaining $\lambda_{i}$ become determined by
$\lambda_{i+1}=\lambda_{i-1}$. Once $\lambda$ is fixed the requirement that
all polynomials $p_{i}(y)$ be normal, determine the coefficients
$a_{i},\tilde{a}_{i}$ up to $l\text{th-}$ roots of unity,
$l=l_{1}\,\cdots\,l_{k-1}\,(l_{k}\,l_{k+1}-1)$. Additionally we have
\begin{equation*}
    \hat{\epsilon}_{i}=\lambda_{i}\,\epsilon_{i},\quad
    \hat{\delta}_{i}=\lambda_{i-1}\,\delta_{i},\quad
    \hat{\epsilon}_{i}^{2}=\hat{\omega}_{i},\quad
    \hat{\delta}_{i}^{2}=\hat{\omega}_{i-1},\quad i=1,k+1.
\end{equation*}
In that case $\hat{\omega}_{1}=\lambda_{1}^{2}\,\omega_{1}$. Therefore the
possible $\hat{\omega}_{1}$ are of the form $\zeta\,\omega_{1}$, with
$\zeta\in\calRE^{2}$. Respect to which $\hat{\omega}$ are allowed, there follows
similar conclusions to those obtained in the case of normal form \Raa, after
replacing $\calN$ with $\calRE$ and $\mathcal{N'}$ with $\calRE^{2}$.

Finally, the last assertion of the Theorem follows by direct calculation.
\end{proof}
\begin{cor}\label{cor:reversors}
A map $f\in \calA\cup\calE\setminus\calS$ is a reversing symmetry for some
nontrivial, cyclically reduced automorphism $g\in\calG$ if and only if $f$ is
$\calS$-conjugate to either a normalized affine reversor or to a normalized
elementary reversor.
\end{cor}
\begin{cor}
Every polynomial involution is conjugate to either of the normal
involutions,
\begin{inparaenum}[\itshape i)\upshape]
    \item $(x,y)\rightarrow (p(y)-x,y),\ p(y)$ a normal polynomial,
    \item $(x,y)\rightarrow (p(y)-x,-y)$,\ $p(y)$ normal and even,
    \item $(x,y)\rightarrow (p(y)+x,-y)$,\ $p(y)$ normal and odd, or
    \item $(x,y)\rightarrow (y,x)$.
\end{inparaenum}
\end{cor}
\begin{proof}Every involution is a reversing symmetry for some nontrivial
automorphism.
\end{proof}
\begin{cor}
An elementary, nonaffine, map
\begin{equation}\label{eq:general-e}
    e:(x,y)\rightarrow (p(y)-\delta\,x,\epsilon\,y+\eta)
\end{equation}
is a reversing symmetry of a nontrivial, cyclically reduced map in $\calG$ if
and only if it has finite even order $2n$, $e^{2}\in\calS$, and $\epsilon^{2},\delta^{2}$
are primitive $n^{th}$-roots of unity.

On the other hand an affine, nonelementary, map
\begin{equation}
a:(x,y)\rightarrow \hat{a} (x,y)+(\xi,\eta),
\end{equation}
$\hat{a}$ a linear transformation, is a reversing symmetry of
a nontrivial, cyclically reduced map in $\calG$ if and
only if it has finite even order $2n$, and $a^{2}\in\calS$.
\end{cor}
\begin{proof}
That these conditions are necessary follows easily from \Cor{cor:reversors}. To
see the sufficiency we prove that elementary (resp.\ affine) maps
satisfying such conditions are $\calS$-conjugate to normalized elementary
reversors (resp.\ normalized affine reversors).

Let us consider the case of an elementary map \Eq{eq:general-e} of order $2n$,
such that $\epsilon^{2},\delta^{2}$ are primitive $n^{th}$-roots of unity and
$e^{2}$ is an affine transformation.
It is not difficult to see that the condition $e^{2n}=id$
implies that if $\eta\ne 0$ then $\epsilon\ne 1$. This observation allows to
prove that $e$ can always be conjugated to an elementary map \Eq{eq:general-e},
having $\eta=0$. Moreover, the conjugating maps can be chosen in $C_{\calS}(t)$.

Next, we see that when $\eta=0$ the conditions $e^{2}\in\calS,\ e^{2n}=id$
reduce to the fact that $\epsilon^{2n}=\delta^{2n}=1$, plus the existence of
some constants $A$ and $B$ such that,
\begin{align*}
    p(\epsilon\,y)-\delta\,p(y)&=A\,y+B,\\
    A(\delta^{2n-2}+\delta^{2n-4}\,\epsilon^{2}+\dots+\epsilon^{2n-2})&=0,\\
    B(\delta^{2n-2}+\delta^{2n-4}+\dots+1)&=0.
\end{align*}
Straightforward calculations then show that it is possible to choose maps in
$\calS$ that conjugate $e$ to a normalized elementary reversor. It may be
interesting to note that if $\epsilon^{2}\ne\delta^{2}$ the condition
$e^{2n}=id$ may be omitted and still it can be granted that $e$ is an
elementary reversor of order $2n$.

The case of affine reversors can be worked out in a similar way. It is
convenient to prove first that, under the given conditions, an affine map is
$\calS$-conjugate to its linear part. To obtain this result it is useful to
treat the cases $n=1$ and $n\ge 2$ separately. If $n\ge 2$, the
conditions $a^{2n}=id,\ a^{2}\in\calS$ ($a$ nonelementary) are equivalent to
$\hat{a}^{2}=\omega\,(id)$, where $\omega=-\det \hat{a}$ is a root of unity of
order $n$. When $n=1$ we need, in addition, that the vector $(\xi,\eta)$
be an eigenvector of $\hat{a}$ with associated eigenvalue $-1$. Finally, it is not
difficult to check that a linear, nonelementary map $\hat{a}$ of order $2n$,
that satisfies the hypothesis $\hat{a}^{2}\in \calS$, is $\calS$-conjugate to
$\tau_{\omega}$ for $\omega=-\det\hat{a}$.
\end{proof}
\begin{cor}
A polynomial automorphism is reversible by involutions in $\calG$ if and
only if it is conjugate to any of the normal forms \Raa, \Rae, or \Ree, with
$\omega=1$, so that $e_{1}$ and $e_{k+1}$ are normal elementary involutions and
$h_{i},\ i=1,\dots,k$, are arbitrary normal \Henon transformations.
\end{cor}
\begin{cor}
A real polynomial automorphism has real reversors in $\calG$ if and only if it
is reversible by involutions, so that it is conjugate to one of the (real)
normal forms \Raa, \Rae, or \Ree, with $\omega=1$, or if it has a reversing
symmetry of order $4$ and is conjugate to a normal form map \Raa, with
$\omega=-1$, so that the maps $h_{i}$ represent normal, real, \Henon
transformations, whose respective polynomials $p_{i}(y)$ are odd.
\end{cor}
\begin{proof}
We noted earlier that the only possible real, reversing symmetries are of order
$2$ or $4$. However there are no elementary, real, reversing symmetries of
order $4$, since this would imply that $e$ is of the form \Eq{eq:general-e},
with $\epsilon^{2},\delta^{2}$ primitive square roots of unity. Therefore the
only possible normal form for a real map with a real reversor of order $4$ is
\Raa, with $\omega=-1$.
\end{proof}

\section{Examples} \label{sec:examples}
In this section we illustrate some of the concepts and results of
\Sec{sec:symmetries} and \Sec{sec:reversors}. We present several
examples to illustrate the general theory. We do not assert that
these examples are necessarily new, merely illustrative---examples of maps with nontrivial symmetry
groups are well known \cite{Zaslavsky91,Golubitsky02}. Examples of
maps with noninvolutory reversors have been presented before; for
example, Lamb found ``modified Townsville" with
a reversor of order $4n+2$ for any $n$---these maps also
have involutory reversors \cite{Lamb92}.  In addition, Roberts and
Baake have shown that ``generalized standard maps," a particular
case of semilength-two maps, can have $4^{th}$ order reversors
\cite{Roberts03}.

For the case of semilength-one, i.e. a single generalized \Henon
map, \Eq{eq:henon}, the structure of the reversing symmetry group is well known. 
As shown in \Prop{thm:affineSym}, there is an
affine-elementary symmetry $S(x,y) = (\omega x ,\omega y)$ for any
$\omega \in \calRA = \{\omega : p(\omega y) = \omega p(y)\}$. Since
we can take $\omega$ to generate $\calRA$, and $S$ and $h$ commute,
$\Sym{h} = \Grp{h} \times \Grp{S}\simeq \Z \times \calRA$.
For example when $p$ is odd, the \Henon
map has the reflection symmetry $S(x,y) = (-x,-y)$.

The reversible cases of \Eq{eq:henon} can be obtained using
\Eq{eq:deltas}. When $\delta \ne \pm 1$, there are no reversors,
so that $\Rev{h} = \Sym{h}$. If $\delta = 1$, then $h$ has the
involutory reversor $t$, which commutes with $S$ so that $\Rev{h} 
    = \Sym{h} \rtimes \Grp{t} = (\Grp{h} \rtimes \Grp{t}) \times \Grp{S}$. 
The case $\delta = -1$ is reversible with reversor $R =
(\epsilon y, \epsilon x)$, providing $\epsilon^2 \in \calRA$ and
$p(\epsilon y) = -\epsilon p(y)$. There are two possibilities: if
the order of $\calRA$ is odd, then there is an involutory reversor
$R(x,y) = -(y,x)$, and the group has the same structure as the
case $\delta = 1$ (for the real case this is the same as Table 5
of \cite{Roberts03}). Otherwise the reversors are complex and
noninvolutory. In this case there is an $\epsilon$ such that the group $\Grp{R}
\simeq \Z_{2k}$ gives all affine-elementary symmetries
and reversors, and thus we can write $\Rev{h} = \Grp{h} \rtimes
\Grp{R} \simeq \Z \rtimes \Z_{2k}$.

Thus we conclude that if $k$ is the order of $\calRA$, then the symmetries of the generalized \Henon map are
\begin{enumerate}
   \item $\delta \neq \pm  1 \quad \Rightarrow \quad \Rev{h} = \Sym{h} = \Grp{h} \times \Grp{S}$,
   \item $\delta = 1$ or $\delta = -1$ and $k$ is odd  $\Rightarrow \quad \Rev{h} =  (\Grp{h} \rtimes \Grp{t}) \times \Grp{S}$,
   \item $\delta = -1$ and $k$ is even $ \Rightarrow \quad \Rev{h} = \Grp{h} \rtimes \Grp{R}$,
\end{enumerate}
where $\Grp{h} \simeq \Z$, $\Grp{S} \simeq \calRA \simeq \Z_k$, $\Grp{t} \simeq \Z_2$, and $\Grp{R} \simeq \Z_{2k}$.
  
Similarly the symmetries for the
semilength two case, $g = h_2 \, h_1$, are also easily found; the
results are given in \Tbl{tbl:semi2}. There are two possible forms
\Se\  and \Sa, corresponding to the groups $\calRE$ and $\calRA$,
respectively. For example in case \Se, there is a symmetry $s(x,y)
=(\zeta(\omega)x/\omega , \omega y)$ for any $\omega \in \calRE$.
If $\calRA$ is trivial or $\delta_1 \ne \delta_2$, then there are
no other nontrivial symmetries, so that $\Sym{g} = \Z \times
\calRE$. However, when $\delta_1 = \delta_2$, then there can be
additional nonaffine symmetries corresponding to case \Sa\
providing the polynomials $p_1$ and $p_2$ are related by the
scaling shown in the table.

According to \Th{thm:normalforms},  there are also two possible
reversible cases of semilength-two, corresponding to normal forms
\Raa\ and \Ree. The conditions for the existence of these can by
found by using \Eq{eq:deltas}; they are also given in
\Tbl{tbl:semi2}. Thus, for example, when there is a normal form
\Raa, the polynomials in $h_1$ and $h_2$ must be identical up to a
scaling. Moreover, if there are noninvolutory reversors, then the
group $\calRA$ must be nontrivial, which implies that
$p_{i}(y)=y\,q_{i}(y)$ for some polynomials $q_{i}(y)$ such that
the degrees of their nonzero terms are not coprime.

Whenever a reversible map has a noninvolutory reversor, then it
also has nontrivial symmetries. For example, for the case \Raa\ in
\Tbl{tbl:semi2}, a noninvolutory reversor corresponds to $\omega
\in \calRA \setminus \{1\}$. In this case $\tau_\omega^2 = s$ is a
symmetry, since $\omega \in  \calRA \subset \calRE$. If in
addition $\delta_1 = \delta_2=1$, then $e^{-1} = e$, and the map has
a ``square root", and  consequently a symmetry of the form
$\tilde{s} h$.

\renewcommand{\arraystretch}{1.5}
\begin{table}[ht]
\begin{center}
\begin{tabular}{|c|c|c|c|l|}
\hline
    Case &
    Normal Form &
    Symmetries &
    Conditions on $\delta_{i}$&
    Conditions on $p_{i}(y)$ \\
\hline\hline
    \multicolumn{5}{|c|} { $\omega\in \calRE(p_{1}(y),p_{2}(y))$} \\
\hline
    \Se &
    $h_2\,h_1$ &
    $s$ &
    {\em arbitrary} & \\
\hline
    \raisebox{-3ex}[0pt]{\Ree} &
    \raisebox{-3ex}[0pt]{$\hat{e}_1^{-1}\, t \, \hat{e}_2 \, t$} &
    \raisebox{-3ex}[0pt]{$\hat{e}_1, \, \hat{e}_2$} &
    \raisebox{-3ex}[0pt]{$\delta_{1}^{2}=\delta_{2}^{2}=1$}&
    \raisebox{-1ex}[0pt] {$p_{1}(\hat{\epsilon}_{1}\,y)= \delta_{1}\,
            \hat{\epsilon}_{2}\,p_{1}(y),\ \hat{\epsilon}_{1}^{2} = \omega$}\\
    & & & & \raisebox{1ex}[0pt]{$p_{2}(\hat{\epsilon}_{2}\,y)=
            \delta_{2}\,\hat{\epsilon}_{1}\,p_{2}(y)$}\\
\hline \hline
\multicolumn{5}{|c|} { $\omega\in \calRA(p_{1}(y),p_{2}(y))$} \\

\hline
    \Sa &
    $s(\tilde{s} h)^2$ &
    $s,\, \tilde{s}h$ &
    $\delta_1 = \delta_2$ &
    $ cp_2(cy) = \omega p_1(y)$\\
\hline
    \Raa &
    $\tau_\omega^{-1}\, h^{-1}\,\tau_\omega h$ &
    $\tau_\omega$ &
    $\delta_{1}\,\delta_{2}=1$ &
    $c\,p_{2}(c\,y)=\delta_{2}\,\omega\,p_{1}(y)$\\
\hline
\end{tabular}
\renewcommand{\arraystretch}{1.0}
\caption{\label{tbl:semi2} Conditions for a semilength-two \Henon
normal form map, $g=h_{2}\,h_{1}$, to have symmetries or be
reversible. Here the symmetry $s(x,y) =
(\frac{\zeta(\omega)}{\omega} x, \omega y)$ is order $k$, and the
reversors $\tau_\omega$ and $e_i$ are order $2k$, where $k$ is the
order of $\omega$.}
\end{center}
\end{table}
    
    For longer words the degrees of the terms can be useful as an indication of
which terms may be centers of symmetry, since the normal forms in
\Th{thm:normalforms} have polydegrees that are symmetric about the centers,
and the polydegree of a reduced word is a conjugacy invariant. However,
explicit conditions analogous to those in
\Tbl{tbl:semi2} are more difficult to write.
    
 Finally we give several examples to illustrate \Tbl{tbl:semi2}.
 \begin{ex}
As an example  consider the composition of two cubic \Henon maps,
$g = h_2 h_1$ where
\begin{equation}\label{eq:cubicHenon}
    h_{1}(x,y)=\left(y,y^3+y -\delta_1x\right),\quad
    h_{2}(x,y)=\left(y,y^3-y-\delta_2 x\right),
\end{equation}
For $(p_1,p_2)$,
\[
     \calN' = \calRE^2 =\{1\} \subset \calRA = \calRE =\calU_{2} \;;
\]
Since $\calRE$ is nontrivial, this system has symmetries of the
form \Se; indeed, since $\zeta(\omega) = 1$, the affine
transformation $S_1(x,y) = (-x, -y)$ is a symmetry. If $\delta_1
\ne \delta_2$, all symmetries are of the form $S_1^a\,g^p$ so
$\Sym{g} = \Grp{g} \times \Grp{S_1} \simeq \Z \times \Z_2$. If
however $\delta_1 = \delta_2 = \delta$, then since $ip_2(iy) =
p_1(y)$, there are symmetries of the form \Sa. We find that $g =
(S_2)^2$ with the symmetry $S_2 = (-i y, i p_1(y) - i\delta x)$.
As the group generated by $S_2$ is still isomorphic to $\Z$, we
have $\Sym{g} = \Grp{S_2} \times \Grp{S_1} \simeq \Z \times \Z_2$.

There are two possible reversible cases for $g$. When $\delta_i^2
= 1$, \Tbl{tbl:semi2} shows that $g$ potentially can be put in
normal form \Ree. The scaling relations imply that
$\delta_1=\delta_2 = \delta=\pm 1$ for this to be the case. Then
$R_1 = (p_1(y)-\delta x, \delta y)$ is an involutory reversor, and
generates a family \Eq{eq:family} of reversors that contains
all reversors with this ordering.

When $\delta_1 \delta_2 = 1$, \Eq{eq:cubicHenon} potentially has a
reversor with normal form \Raa. In this case the scaling relations
also require $\delta_1 = \delta_2 = \delta = \pm 1$, and there is
a reversor $R_2 = (-i\delta y,i x)$. When $\delta = 1$,  $R_2$ is
an involution; however, when $\delta = -1$, it is order four and
$R_2^2 = S_1$. In both cases, $\delta = \pm 1$, there is an
involutory reversor,$R_1$; therefore, every reversor can be
written as the composition of a symmetry and $R_1$; for example
$R_2 = S_2\,R_1$.

Thus we conclude that there are three distinct cases:
\begin{enumerate}
  \item $\delta_1 = \delta_2 \in \calU_2 \;\Rightarrow\;
     \Rev{g} = (\Grp{S_2} \times \Grp{S_1}) \rtimes \Grp{R_1}$,
   \item $\delta_1 = \delta_2 \not\in \calU_2 \;\Rightarrow\;
     \Rev{g} = \Sym{g} =\Grp{S_2} \times \Grp{S_1}$,
  \item $\delta_1 \ne \delta_2 \;\Rightarrow\;
     \Rev{g} = \Sym{g} = \Grp{g} \times \Grp{S_1}$,
\end{enumerate}
where $\Grp{S_2}\simeq \Z$, $\Grp{S_1}\simeq \Z_2$ and $\Grp{R_1} \simeq \Z_2$.
\end{ex}

\begin{ex}\label{ex:noreal}
Let $g = h_2 h_1$ where
\begin{equation*}
    h_{1}(x,y)=\left(y,y^3+x\right),\qquad
    h_{2}(x,y)=\left(y,y^3-x\right),
\end{equation*}
In this case the associated groups of roots of unity are larger:
\[
     \calN' =\calRA = \calU_2 \subset \calRE^2 =
       \calU_4 \subset \calN = \calRE = \calU_8 \;.
\]
and  $\zeta(\omega) = \omega^4$. \Tbl{tbl:semi2} shows that
the nontrivial symmetries are generated by $S_1 = (\omega^3 x, \omega y)$
with $\omega = e^{i \frac{\pi}{4}}$ a primitive, eighth root of unity, and
$\Grp{S_1} \simeq \calU_8$. Since $\delta_1 \neq \delta_2$ there are no symmetries of the form \Sa. Thus $\Sym{g} = \Grp{g}\times \Grp{S_1} \simeq \Z \times \Z_8$.

Since $\delta_1 \delta_2 \ne 1$, $g$ cannot be written in form
\Raa; however, it can be written in form \Ree, for
$\hat{\epsilon}_{1}^{8} = -1$. The reversors are generated by $R_1
= (\hat{\epsilon}_1(y^3+x), \hat{\epsilon}_{1}^3 y),$ a sixteenth
order reversor. Note that $R_1^2 = S_1$, and that $R_1$ commutes
with $S_1$. Thus $\Rev{g} = \Grp{g} \rtimes \Grp{R_1}  \simeq (\Z \rtimes Z_{16})$.

Note that $\calRE^{2}=\calU_{4}$,\ so that
$\omega$ in the \Raa\ normal form may be replaced only by primitive
$8$th-roots of unity. Thus, there are no real normal forms and though $g$ is reversible in the group of complex automorphisms, it lacks real reversors.
\end{ex}

\begin{ex}\label{ex:nocomm}
Consider the $g=h^2$, where $h$ is a normal
\Henon transformation. Symmetries of the form \Se\ correspond to maps 
$S^{j}\,g^{p}$ with $S$ a generator of the group of
affine-elementary symmetries of $g,$ that is $S=s_{\omega}$ for
$\omega$ of maximum order in $\calRE$.

Symmetries of the form \Sa\ correspond to maps  $s\,\tilde{g}^{p}$, where $s$ is an
affine-elementary symmetry of $g$ and $\tilde{g}=\tilde{s}\,h$ is
a symmetry of $g$ that commutes with $s$. It turns out
that $s$ is also in $\Sym{h},$ so that $s=s_{\omega}$ for some
$\omega\in\calRA$. Moreover, since $h\in\Sym{g},$ it follows
that $\tilde{s}$ also belongs to $\Sym{g}$, so that
$\tilde{s}=s_{\omega}$ for some $\omega\in\calRE$. Thus we can
conclude that $\Sym{g}=\Grp{S,h}$. If $\calRA = \calRE$, then $S$ is a symmetry of $h$, and so $\Sym{g} =\Grp{S} \times \Grp{h} \simeq \Z_k \times \Z$.
On the other hand if $\calRA\ne\calRE$, $S$ is not a symmetry
of $h$ and, unlike the previous examples, $\Sym{g}$ is
a nonabelian group. In this case, however, $\Grp{S}$ is a normal subgroup of $\Sym{g}$ so that $\Sym{g}=\Grp{S}\rtimes\Grp{h} \simeq \Z_k \rtimes \Z$.

According to \Tbl{tbl:semi2}, the existence of reversors of the form \Ree\
requires $\delta^{2}=1$ plus some scaling conditions on the
polynomial $p(y)$. The associated reversors are in that case of
the form
\[
     R(x,y)= (\frac{1}{\hat{\epsilon_2}}(p(y)-\delta\,x),
               \frac{1}{\hat{\epsilon_1}}\,y),
\]
$\epsilon_1,\epsilon_2$ as in \Tbl{tbl:semi2}. For $\delta=1$ we
see that the scaling conditions are trivially satisfied when
$\hat{\epsilon}_1=\hat{\epsilon}_2=1$ so that $R$ is an involution. On the other hand, when
$\delta=-1$ the scaling conditions are satisfied only when $p(y)$
is odd or even. In that case ${\hat{\epsilon_i}} \in \{\pm 1\}$ and again
$R$ is an involution.

Reversors of the form \Raa\ exist only if $\delta^2=1$ and $p(y)$ satisfies
the condition $c\,p(c\,y)=\delta\,\omega\,p(y)$ with $\omega\in\calRA$ for some constant
$c$. The associated affine
reversing symmetry is then of the form
$R(x,y)=(cy/\!\omega,x/\!c)$. Again it can be seen that when
$\delta=1$ the scaling relation is trivially satisfied taking
$c=\omega=1$, while when $\delta=-1$ that relation is satisfied if
and only if $p(y)$ is either an odd or an even polynomial.

We thus have the following two possible structures for the group
of reversing symmetries of $g$:
\begin{enumerate}
  \item $\delta\neq \pm 1$ or $\delta=-1$ with $p(-y)\neq\pm p(y)$ $\Rightarrow$
        $   \Rev{g} = \Sym{g} = \Grp{S}\rtimes\Grp{h} $,
  \item $\delta=1$ or $\delta=-1$ {with} $p(-y)=\pm p(y)$  $\Rightarrow$
       $ \Rev{g} =  (\Grp{S}\rtimes \Grp{h})\rtimes\Grp{R}$,
\end{enumerate}
where $ \Grp{S} \simeq \Z_k$, $\Grp{h} \simeq \Z$, and $\Grp{R} \simeq \Z_2$ and  $k$ is the order of $\calRE$. If $k$ is also the order of $\calRA$, then $\Sym{g} = \Grp{S} \times \Grp{h}$.

\end{ex}
\section{Dynamics}

The dynamics of a map is affected in a number of ways by the existence
of reversing symmetries.  In particular, those orbits that are
``symmetric'' share many of the typical properties of the orbits of
symplectic maps, i.e., their spectral and bifurcation properties.
Although our main interest is to discuss plane polynomial
diffeomorphisms, we begin with a general discussion, and later focus
on the polynomial case.  We start by briefly reviewing some of the well known
implications of reversibility \cite{Roberts92b, Lamb98a}.

Let $\calO(x)$ denote the orbit of $x$ under a diffeomorphism $g$.
When  $R \in \Rev{g}$, then the symmetry maps
orbits into orbits, $R(\calO(x))=\calO(R(x))$. Thus orbits either
come in symmetric pairs, or are themselves invariant under $R$.  If $R$ is a reversor, then an orbit and
its reflection are generated in reverse order.

If $\calO(R(x))=\calO(x)$, the orbit is said to be {\em symmetric} with
respect to $R$.  Observe then that the orbit
is symmetric respect to any of the reversing symmetries in the subgroup
$\Grp{g,R}$ generated by $g$ and $R$.

If $R$ is a reversor, then for symmetric orbits
forward stability implies backward stability. Moreover, there can be no
attractors that are symmetric under $R$; indeed,
if $A$ is a symmetric omega-limit set, then
it cannot be asymptotically stable \cite{Lamb98c}.

By contrast, an asymmetric orbit can be attracting, just as long as its
symmetric partner is repelling.

If $R$ is a reversor, and $x$ is a point on a symmetric orbit of period $n$, then the matrix $Dg^n(x)$
is conjugate to its inverse.  Thus every multiplier of a periodic
symmetric orbit must be accompanied by its reciprocal.  When $g$ is
real, it follows that eigenvalues other than $\pm 1$ must appear
either in pairs $(\lambda,\lambda^{-1})$, with $\lambda$ real or on
the unit circle, or in quadruplets,
$(\lambda,\bar{\lambda},\lambda^{-1},\bar{\lambda}^{-1})$.  Unlike the
symplectic case, $1$ and $-1$ may have odd multiplicity.  This
situation imposes severe restrictions on the motion of eigenvalues for
parameterized families of maps: thus if the multiplicity of $1$ or
$-1$ is odd, it must continue to remain so, as long as reversibility is
preserved.  Therefore, whenever $1$ or $-1$ have odd multiplicity,
they should persist as eigenvalues.  In the plane this means that for
families of reversible, orientation-reversing maps, the spectrum is
restricted to the set $\{1,-1\}$.  In all the other cases families of
reversible maps must preserve orientation.

Though generally reversible maps need not be volume-preserving,
reversible polynomial automorphisms are, since their jacobians are necessarily constant.
In addition, note that maps with the normal form
\Raa\ are orientation-preserving. Maps with the normal form \Rae\
or \Ree\ can either preserve or reverse orientation.

We denote the fixed set of a map $R$ by
\[
    \Fix{R} \equiv \{x: R(x) = x\} \;.
\]
If $S \in \Sym{g}$ then its fixed set is an invariant set. By contrast, the fixed sets of reversors
are not invariant, but contain points on symmetric orbits.
%

%
%
Indeed, as is well-known, to look for symmetric periodic orbits
it is enough to restrict the search to the set
$\Fix{R}\cup\Fix{g\,R}$ \cite{Devaney76,Roberts92b,MacKay93,Lamb98b}.
Therefore if the reversing symmetry has a
nontrivial fixed set, it can be used to simplify the computation of
periodic points.  Indeed, Devaney's original definition of
reversibility \cite{Devaney76} required that the fixed set of the
reversor be a manifold with half the dimension of the phase space.
This is the case for maps on the plane that are reversible by
orientation-reversing involutions \cite{MacKay93}.  From this point of
view the noninvolutory polynomial reversors we have described are not
very interesting, since for each of them the associated symmetric
orbits reduce to a single fixed point.

We will now show that this is always the case for order $4$ reversing symmetries
of the plane. In addition, we will recall the result that in this case the
symmetric fixed point is hyperbolic \cite{Lamb93}.

We start by showing that the fixed set of any order $4$ transformation
of the plane is a point.  This is a well-known result of Brouwer, who showed that
finite period transformations of $\R^2$ are topologically equivalent
to either a rotation or to the composition of a rotation and a
reflection about a line through the origin \cite{Brouwer19}.  Nevertheless, we
present an elementary proof of the local nature of the fixed set
similar to that given by MacKay for the case of involutions
\cite{MacKay93} because this proof provides additional information
that we find useful later.  To complete the description of the fixed
set some general results on transformation groups turn out to be
necessary.

\begin{lem}
    Suppose $R$ is an order $4$ diffeomorphism of $\R^2$. Then its fixed set is a point.
\end{lem}
\proof Assume first that $R$ has a fixed point, without loss of
generality, at $(0,0)$.  We see that then $f=R^{2}$ is an
orientation-preserving involution, since the jacobian matrix for $f$
at $(0,0)$ equals the square of the corresponding jacobian matrix of
$R$.  A simple calculation then shows that $Df(0,0) = \pm I$.  Thus we
can write
\[
    f:(x,y)\rightarrow \pm (x,y)+(f_{1}(x,y),f_{2}(x,y)),
\]
with $f_{k}(x,y) =o(|(x,y)|)$. Define new variables $u$
and $v$ according to the local diffeomorphism
\[
    (u,v) = \pm(x,y) + \frac{1}{2} \left( f_1(x,y),\, f_2(x,y)\right)
\]
where the sign is chosen in accordance with $Df(0,0)$. In these local variables
the map $f$ reduces to $(u,v)\rightarrow \pm(u,v)$. We thus have shown that at each
fixed point of $R$, $f$ is locally conjugate to $\pm id$. Moreover, whenever $R$
has a fixed point, the above condition holds at each of the fixed points of $f$.
This implies that if $f$ is locally
the identity, then $f=id$ on its domain, as long as this domain is connected.
Therefore, given that $R$ has order $4$ and is not an involution, we conclude
that $(0,0)$ is an isolated fixed point and that the (real) normal
form for $DR(0,0)$ is given by
\begin{equation}\label{eq:jacobian}
    \begin{pmatrix} 0&-1\\1&0 \end{pmatrix}.
\end{equation}

To prove that $R$ actually has a unique fixed point, we proceed to extend the
map $R$ to $S^{2}$, by adding a point at infinity and make it a fixed  point of
$R$. Now, according to Smith's classical results on transformation  groups (see
\cite{Bredon72}), the fixed set for a transformation of order $p^{k}$, ($p$
prime) acting on a $n$-sphere has the homology-$\mod p$ of an $r$-sphere
for some $-1\le r\le n$, where $r=-1$ corresponds to the empty set. We know
however that the fixed set of $R$, acting on the sphere, is nonempty, and that
fixed points (at least other than the fixed point at infinity) are isolated.
Therefore the fixed set of $R$ must have the $\mod p$ homology of $S^{0}$,
hence it consists exactly of two points. Restricted to the plane we see that
$R$ has exactly one fixed point.
\qed

Notice that whenever a reversing symmetry $R$ for some map $g$ has a single
fixed point, this point becomes a symmetric fixed point of $g$. It was shown by Lamb for the
 case that $R$ is a rotation by $\pi/4$ that this point cannot be elliptic \cite{Lamb93}. Using 
 the previous lemma, it is easy to generalize this to arbitrary order 4 reversors.
\begin{lem}\label{lem:hyperbolic}
    Suppose $g$ is a reversible map of $\R^2$ with a real reversor $R$ of order $4$.
    Then the associated symmetric fixed point is not elliptic.
\end{lem}
\proof We may assume that the fixed point is the
origin and the jacobian matrix of $R$ at $(0,0)$ is given by \Eq{eq:jacobian}.
The reversibility condition implies that
\begin{equation*}
DR(0,0)\,Dg(0,0)=Dg(0,0)^{-1}\,DR(0,0).
\end{equation*}
This equation implies that $Dg(0,0)$ is a symmetric matrix with
determinant equal $1$. As symmetric real matrices have real eigenvalues, we
conclude that the point $(0,0)$ cannot be elliptic, and that the map is
orientation-preserving.
\qed

To illustrate some of this phenomena, we give two examples.
\begin{ex}
Consider a normal form of type \Raa:
\begin{align*}
    g     &= \tau_{\omega}^{-1}\,h^{-1}\,\tau_{\omega}\,h  \;, \nonumber\\
    h(x,y)&= (y,y^3-b\,y-\delta\,x) \;,\quad \omega = -1\;,
\end{align*}
so that $\tau_{\omega}(x,y) =(-y,x)$ is an order $4$ reversor for
$g$. If we let $p(y) = y^2-b y$ and assume $b \ne 0$, then $\calRA(p) = \calU_2$, while $\calN' = \{1\}$, so the only reversing symmetries for this ordering are order four.
Fixed points of this map must satisfy the equations
\[
     (1-\delta)x^* = -p(y^*) \;,\quad (1-\delta)y^* = p(x^*) \;
\]
where $p(x) = x^3-bx$. Since the reversor $\tau_\omega$ has a fixed point at the origin,
the origin is always a symmetric fixed point. In general, the stability of a fixed point
is determined by
\[
   \Tr(Dg) = \frac{1}{\delta}\left(p'(y^*)p'(x^*) +\delta^2+1\right)
\]
At the origin this becomes $\Tr(Dg(0,0)) = \delta^{-1}(b^2+\delta^2+1)$, which implies,
in accord with \Lem{lem:hyperbolic}, that
the origin is hyperbolic, since $|\Tr(Dg(0,0)| > 2$, except when
$(b,\delta) = (0,\pm1)$, where it is parabolic.


The remaining $8$ fixed points are born together in $4$ simultaneous
saddle-node bifurcations
when
\[
   b = b_{sn\pm} \equiv \pm 2\sqrt2 |\delta -1| \;.
\]
These lines are shown in \Fig{fig:stability}; inside the cone
$b_{sn-} \le b \le b_{sn+}$ the map $g$ has only one fixed point.
The dynamics of this situation are depicted in \Fig{fig:cubicfig1} for
the case that $\delta = 1.3$.

Outside this cone the map has eight asymmetric fixed points. An
example is shown in \Fig{fig:cubicfig2}; for this case four of the
fixed points are elliptic and four are hyperbolic. Note that the
four islands surrounding the elliptic fixed points in this figure
are mapped into one another by $\tau_\omega$. The elliptic fixed
points undergo a period-doubling when $\Tr(Dg(x^*,y^*)) = -2$,
which corresponds to the curve
\[
 b^4-(7-13\delta+7\delta^2)b^2-2(2\delta-1)^2(\delta-2)^2 = 0
\]
This gives the dashed curves shown in \Fig{fig:stability}.
For example if we fix $\delta=1.3$ and increase $b$, then period-doubling occurs
at $b \approx 1.6792$. The four new period two orbits are stable up to $b
\approx 1.7885$, when they too undergo a period-doubling bifurcation. Thus in
\Fig{fig:cubicfig3}, there are $4$ unstable period-two orbits and $4$ more
corresponding period four orbits.  In this case the stable and unstable
manifolds of the saddles intersect, forming a complex trellis.
\end{ex}

\InsertFig{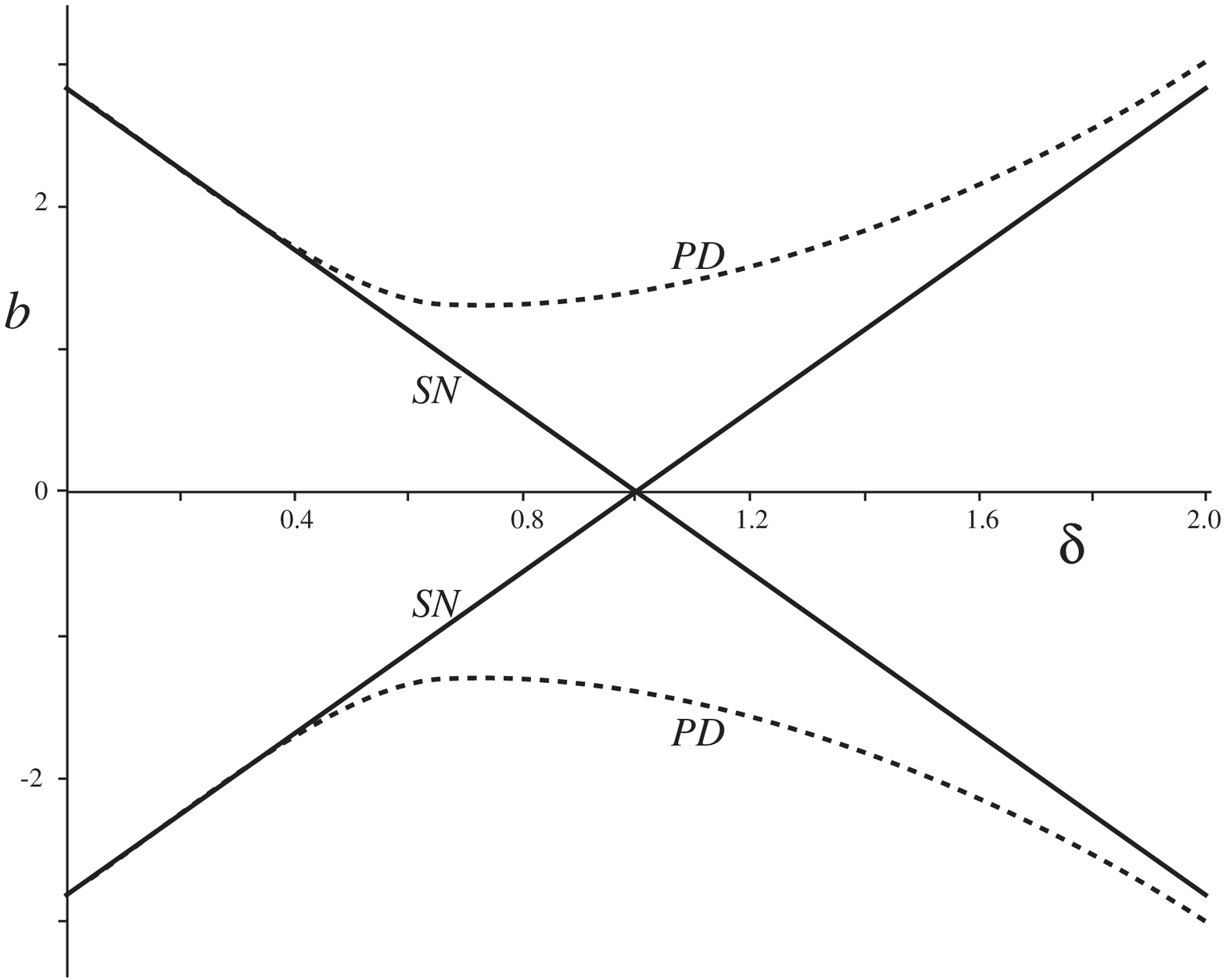} {Bifurcation curves for the asymmetric fixed
points of \Eq{eq:cubicHenon}.  The 8 asymmetric fixed points exist in
the upper and lower quadrants of the cone.  They undergo
period-doubling bifurcations on the dashed curves.}
{fig:stability}{3.0in}

\InsertFig{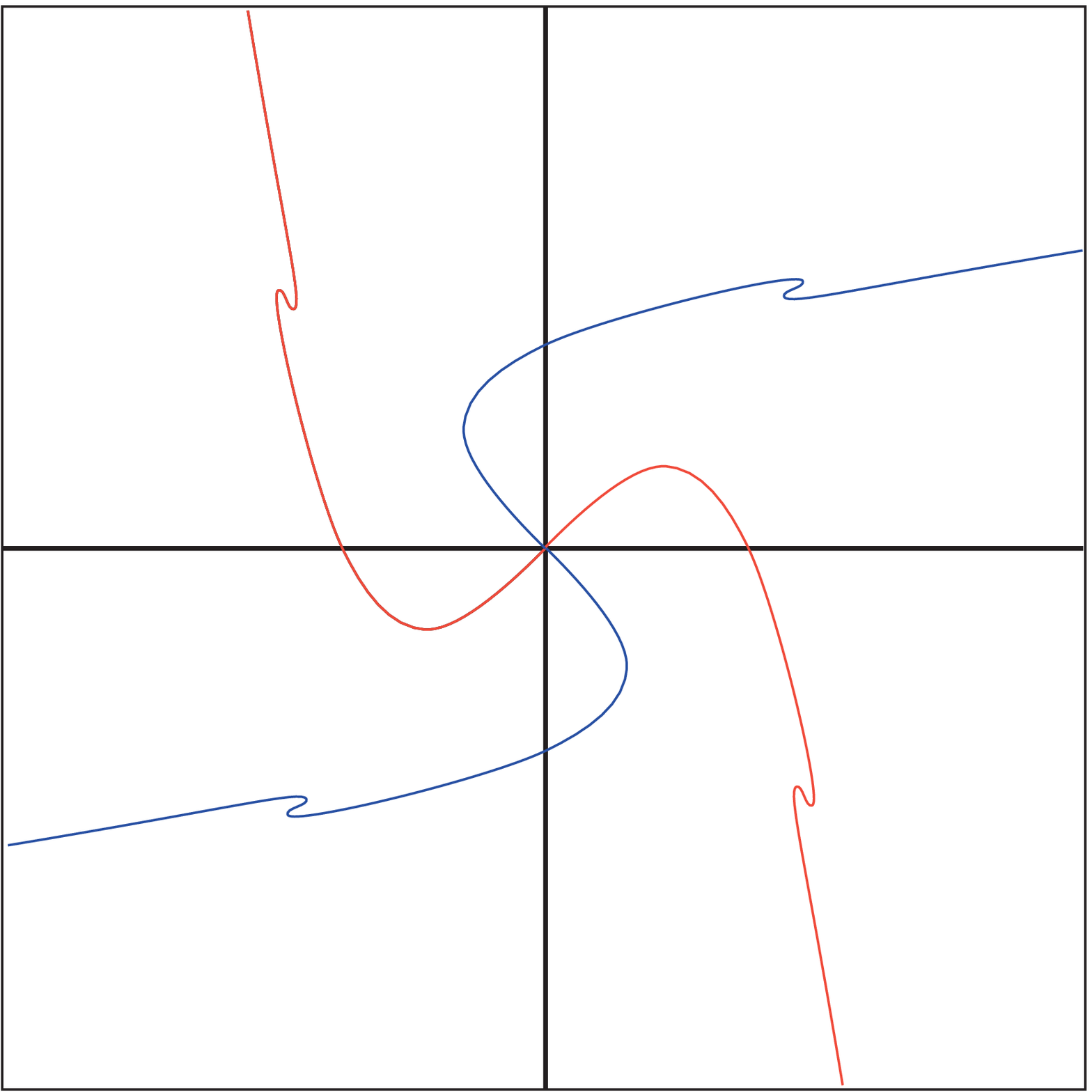} {Stable and unstable manifolds of the origin for
\Eq{eq:cubicHenon} for $(b,\delta) = (0.85,1.3)$.  The domain of the
figure is $(-3,3)\times(-3,3)$.}{fig:cubicfig1}{3.0in}

\InsertFig{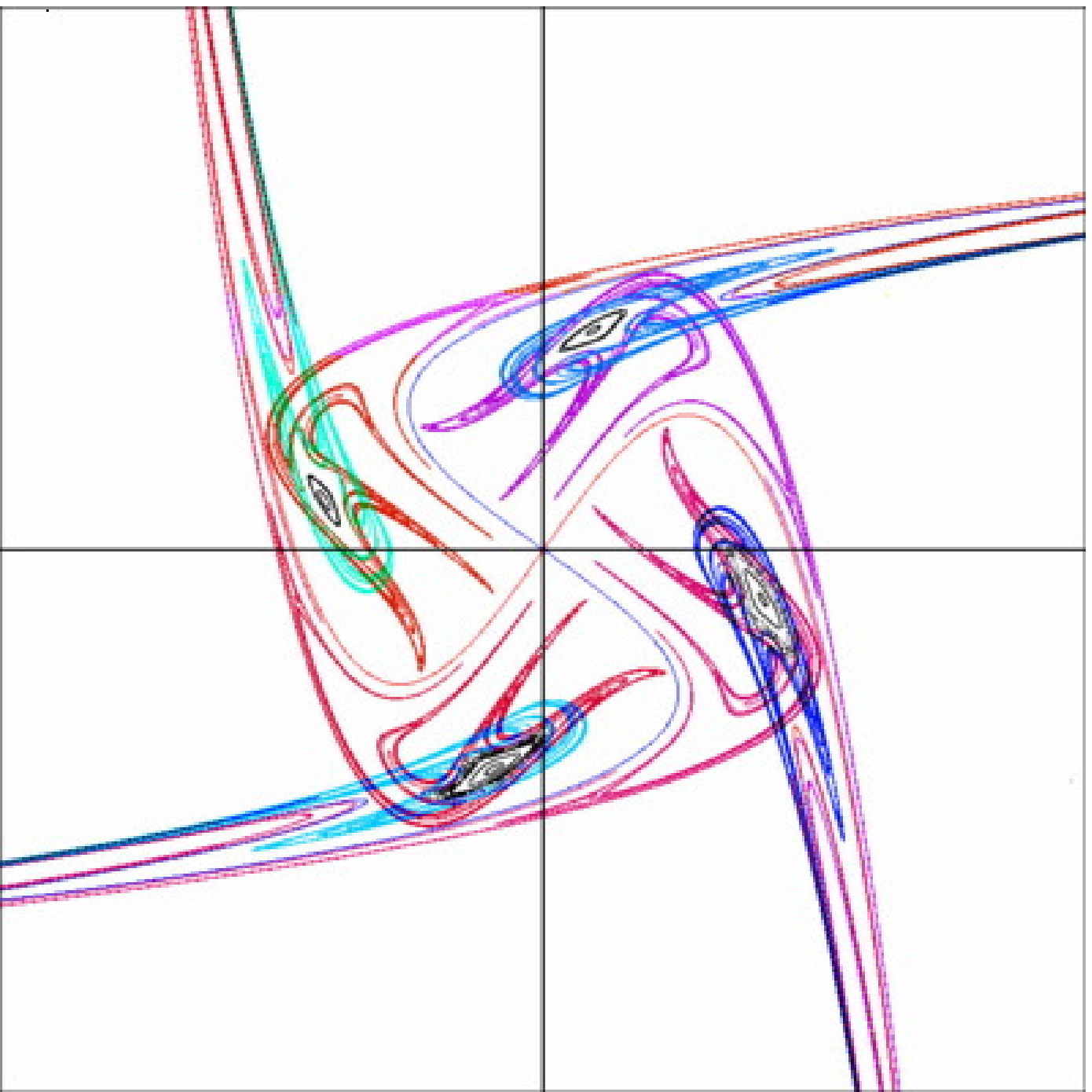} {Some orbits of the map \Eq{eq:cubicHenon} for
$(b,\delta) = (1.4,1.3)$.  There are elliptic fixed points at
$(0.27438,1.2116)$ and hyperbolic points at $(1.0010,1.2794)$, as well
as the images of these points under $\tau_\omega$.  The domain is the
same as \Fig{fig:cubicfig1}.}{fig:cubicfig2}{3.0in}

\InsertFig{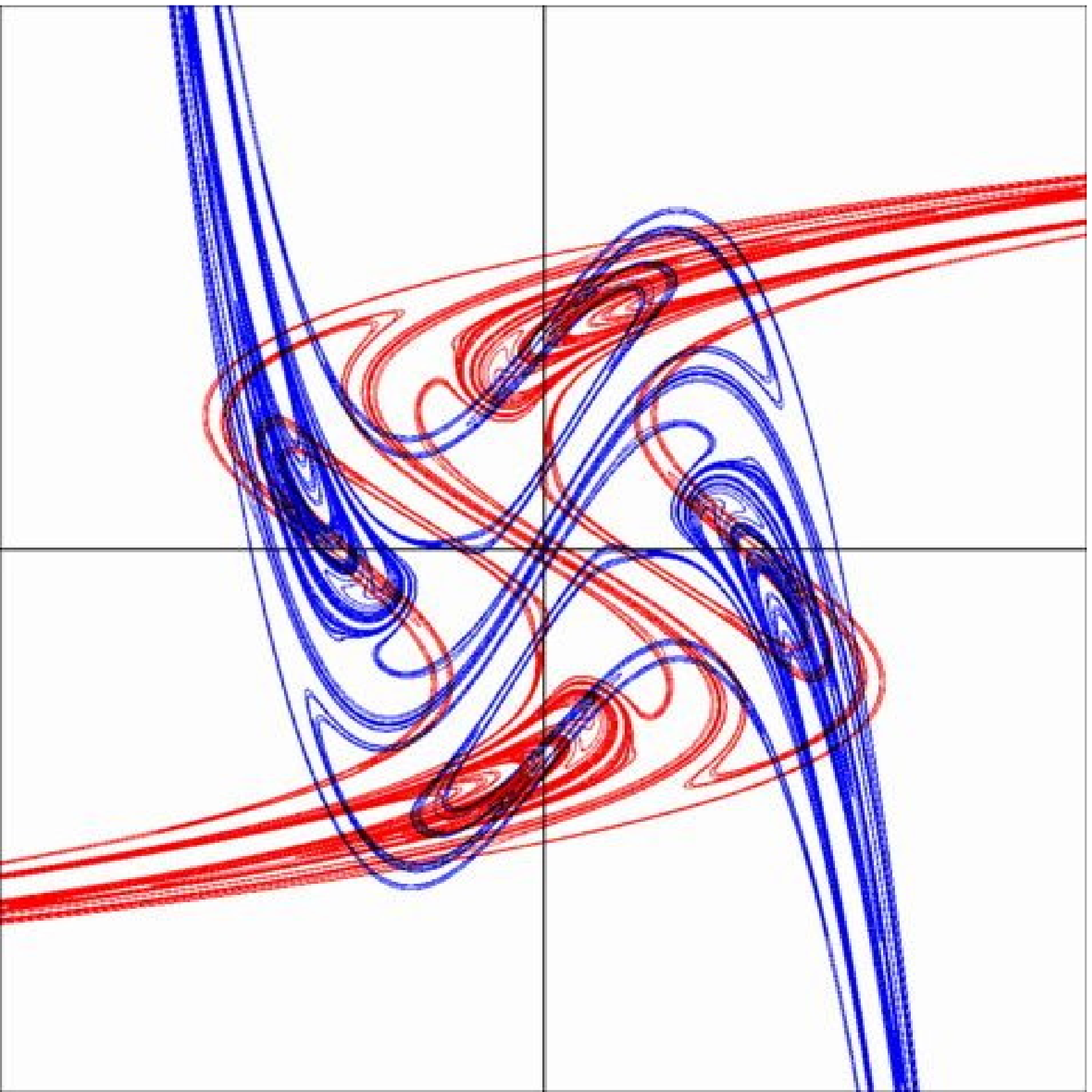} {Some stable and unstable manifolds of the map
\Eq{eq:cubicHenon} for $(b,\delta) = (1.8,1.3)$.  Here the elliptic
points (e.g.~at $(0.23390,1.3607)$) have undergone a period-doubling
bifurcation.  The domain is the same as \Fig{fig:cubicfig1}.}
{fig:cubicfig3}{3.0in}

In contrast to \Lem{lem:hyperbolic}, real maps with order $4$ complex reversors,
can have elliptic
symmetric fixed points.

\begin{ex}
Consider for instance the polynomial map,
given in \Henon normal form
\begin{equation}\label{eq:twoHenons}
     g=h_{2}\,h_{1} \;, \quad\mbox{with} \quad h_{k}(x,y)=(y,p_{k}(y)+x) \;,
\end{equation}
and assume that $i\in\calRA(p_{1}(y),p_{2}(y))$. According to \Tbl{tbl:semi2},
$g$ can be written in normal form \Ree\ with associated order $4$ reversors.
Direct calculations show that $g$ is conjugate to the map
$\hat{g}=t\,\hat{e}_{1}^{-1}\,t\,\hat{e}_{2}$, where
\[
    \hat{e}_{1}(x,y)=(\hat{p}_{1}(y)+i\,x,-i\,y),\quad \text{and}\quad
    \hat{e}_{2}(x,y)=(\hat{p}_{2}(y)-i\,x,i\,y)
\]
are elementary normal reversors and the $\hat{p}_{k}$ are rescalings of
$p_{k}$. Therefore for some scaling $s$ the map
$s^{-1}\,\hat{e}_{2}\,s$ is an order $4$ reversor for $g$ and the origin is a
symmetric fixed point. Furthermore $\Tr(Dg(0,0))=2+p_{1}'(0)\,p_{2}'(0)$, so
that whenever $-4<p_{1}'(0)\,p_{2}'(0)<0$, the origin is an elliptic point.
An example is shown in \Fig{fig:tenthdegree}.

\InsertFig{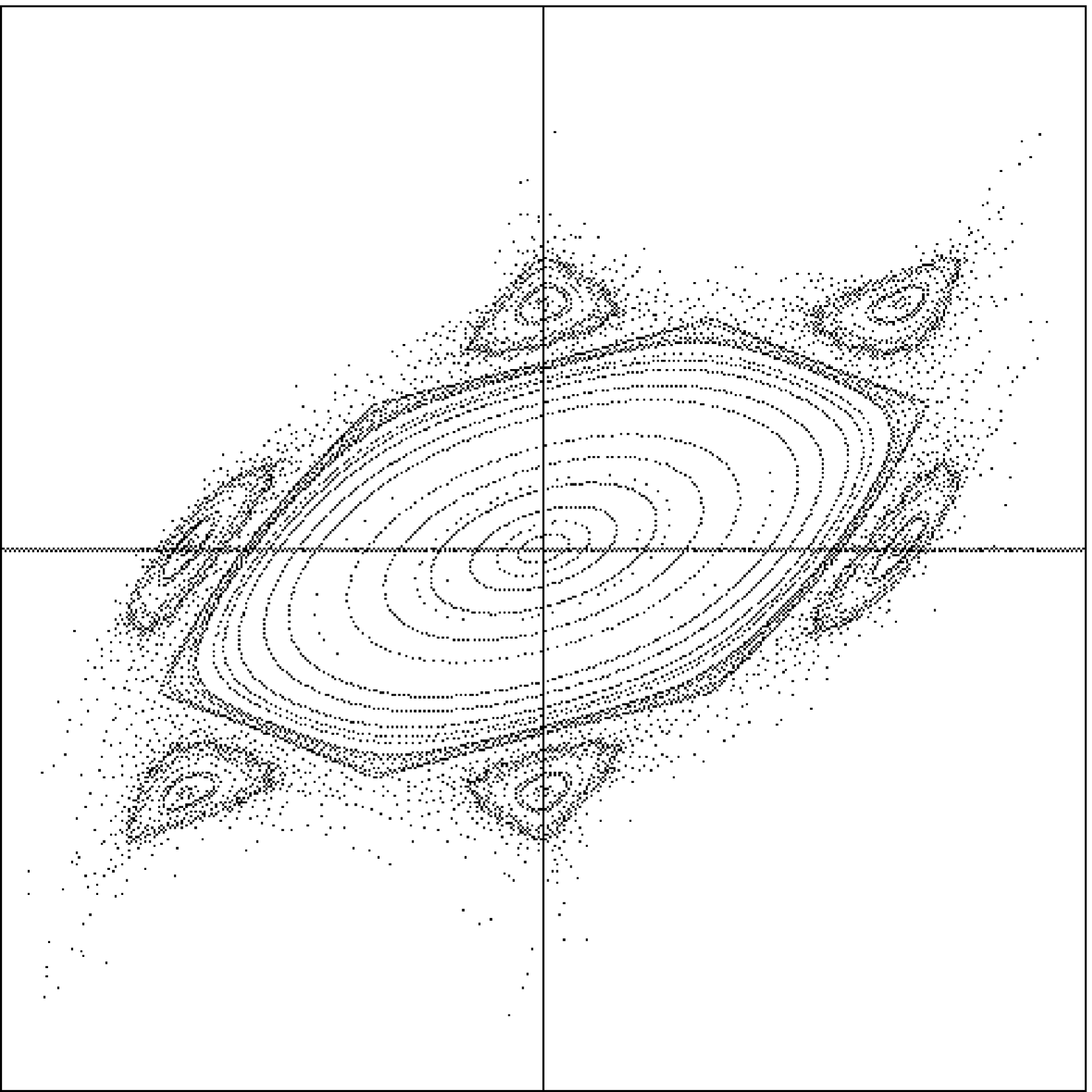}
{A map of the form \Eq{eq:twoHenons} with $p_1 = y^5-0.5y$ and $p_2 = y^5+1.5y$ so that
the fixed point at the origin is elliptic. The domain of the figure is $(-1,1)\times (-1,1)$.}
{fig:tenthdegree}{3.5in}
\end{ex}

In this example the map $g$ also possesses involutory reversors.
In fact this is always the case for orientation-preserving semilength-two maps
with normal form \Ree. That is, whenever the map has order $4$ reversors
there also exist involutory reversors, as can be readily obtained using
conditions on \Tbl{tbl:semi2}.

\section{Conclusions}
We have shown that maps in $\calG$ that have nontrivial symmetries have a normal
form $s_\omega (H)^q$ in which either there is a finite-order, linear symmetry $s_\omega$, or in 
which the map has a root $H$ that is a composition of normal \Henon maps.
The symmetry $s_\omega$ \Eq{eq:normalsymm} generates 
a group isomorphic to $\calRE$ \Eq{eq:RE} if the semilength of the map is even, and $\calRA$ 
\Eq{eq:RA} if it is odd. This result is encapsulated in \Cor{thm:symmetryCorollary}.

Similarly, we have shown that reversors for automorphisms in $\calG$ have normal forms
that are either affine or elementary. These can be further normalized so that
the reversors correspond either to the simple affine map $\tau_\omega$
\Eq{eq:a-reversor}, or to an elementary reversor of the form \Eq{eq:e-reversor}.
These reversors have finite, even order. The case that the order is two, i.e.
involutory reversors, is typical in the sense that the existence
of reversors of higher order requires that the polynomials in the map satisfy extra
conditions so that one of the groups $\calRA$ or $\calRE$
is nontrivial. If a map has real reversors, then they must be order $2$ or $4$.

Using these, we obtained three possible normal forms for reversible polynomial automorphisms of the plane, \Th{thm:normalforms}. These correspond to having either two affine, two elementary,
or one affine and one elementary reversor.

It would be interesting to generalize these results to higher-dimensional polynomial maps.
The main difficulty here is that Jung's decomposition theorem has not been generalized
to this case. Nevertheless, one could study the class of polynomial maps generated by
affine and elementary maps.


\clearpage
\begin{thebibliography}{10}

\bibitem{Henon69}
M.~H\'enon.
\newblock Numerical study of quadratic area-preserving mappings.
\newblock {\em Quart. Appl. Math.}, 27:291--312, 1969.

\bibitem{Henon76}
M.~{H}\'enon.
\newblock A two-dimensional mapping with a strange attractor.
\newblock {\em Comm. Math. Phys}, 50(1):69--77, 1976.

\bibitem{Friedland89}
S.~Friedland and J.~Milnor.
\newblock Dynamical properties of plane polynomial automorphisms.
\newblock {\em Ergod. Th. \& Dyn. Systems}, 9:67--99, 1989.

\bibitem{Jung42}
H.W.E. Jung.
\newblock {\"U}ber ganze birationale {T}ransformationen der {E}bene.
\newblock {\em J. Reine Angew. Math.}, 184:161--174, 1942.

\bibitem{Devaney76}
R.L. Devaney.
\newblock Reversible diffeomorphisms and flows.
\newblock {\em Trans. Am. Math. Soc.}, 218:89--113, 1976.

\bibitem{Sevryuk86}
M.B. Sevryuk.
\newblock {\em Reversible Systems}, volume 1211 of {\em Lecture Notes in
  Mathematics}.
\newblock Springer-Verlag, New York, 1986.

\bibitem{Lamb94b}
J.S.W. Lamb.
\newblock {\em Reversing Symmetries in Dynamical Systems}.
\newblock Ph{D} {T}hesis, Universiteit van Amsterdam, 1994.

\bibitem{Lamb98b}
J.W.S. Lamb and J.A.G. Roberts.
\newblock Time-reversal symmetry in dynamical systems: A survey.
\newblock {\em Physica D}, 112:1--39, 1998.

\bibitem{Lamb92}
J.S.W. Lamb.
\newblock Reversing symmetries in dynamical systems,.
\newblock {\em J. Phys. A.}, 25:925--937, 1992.

\bibitem{Baake97}
M.~Baake and J.A.G. Roberts.
\newblock Reversing symmetry group of ${G}l(2,{Z})$ and ${PG}l(2,{Z})$ matrices
  with connections to cat maps and trace maps.
\newblock {\em J. Phys. A.}, 30:1549--1573, 1997.

\bibitem{Goodson99}
G.R. Goodson.
\newblock Inverse conjugacies and reversing symmetry groups.
\newblock {\em Am. Math. Monthly}, 106:19--26, 1999.

\bibitem{Baake01}
M.~Baake and J.A.G. Roberts.
\newblock Symmetries and reversing symmetries of toral automorphisms.
\newblock {\em Nonlinearity}, 14(4):R1--R24, 1997.

\bibitem{Roberts03}
J.~A.~G. Roberts and M.~Baake.
\newblock Symmetries and reversing symmetries of area preserving mappings in
  generalised standard form.
\newblock {\em Physica A}, 317:95--112, 2003.

\bibitem{Lamb98a}
J.S.W. Lamb, editor.
\newblock {\em Time-Reversal Symmetry in Dynamical Systems}, volume 112 of {\em
  Physica D}, Amsterdam, 1998. Elsevier.

\bibitem{Gomez02}
A.~G{\'o}mez and J.D. Meiss.
\newblock Reversible polynomial automorphisms of the plane: the involutory
  case.
\newblock {\em Physics Letters A}, 312(1--2):49--58, 2003.

\bibitem{Magnus66}
W.~Magnus, Karras A., and D.~Solitar.
\newblock {\em Combinatorial Group Theory}, volume~13 of {\em Pure and Applied
  Mathematics}.
\newblock Interscience Publishers, New York, 1966.

\bibitem{Jacobson85}
N.~Jacobson.
\newblock {\em Basic Algebra}.
\newblock W.H. Freeman, New York, 1985.

\bibitem{Zaslavsky91}
G.M. Zaslavsky, R.Z. Sagdeev, D.A. Usikov, and A.A. Chernikov.
\newblock {\em Weak Chaos and Quasi-Regular Patterns}.
\newblock Cambridge Nonlinear Science Series. Cambridge Univ. Press, Cambridge,
  1991.

\bibitem{Golubitsky02}
M.~Golubitsky and I.~Stewart.
\newblock {\em The Symmetry Perspective : from equilibrium to chaos in phase
  space and physical space}, volume 200 of {\em Progress in mathematics
  (Boston, Mass.)}.
\newblock BirkhÉuser, Basel, 2002.

\bibitem{Roberts92b}
J.A.G. Roberts and G.R.W. Quispel.
\newblock Chaos and time reversal symmetry. {O}rder and chaos in reversible
  dyanmical systems.
\newblock {\em Physics Reports}, 216(2-3):63--1177, 1992.

\bibitem{Lamb98c}
J.S.W. Lamb and M.~Nicol.
\newblock On symmetric attractors in reversible dynamical systems.
\newblock {\em Physica D}, 112(1-2):281--297, 1998.

\bibitem{MacKay93}
R.S. MacKay.
\newblock {\em Renormalisation in Area-Preserving Maps}, volume~6 of {\em
  Advanced Series in Nonlinear Dynamics}.
\newblock World Scientific, Singapore, 1993.

\bibitem{Lamb93}
J.S.W. Lamb.
\newblock Crystallographic symmetries of stochastic webs.
\newblock {\em J. Phys. A.}, 26:2921--2933, 1993.

\bibitem{Brouwer19}
L.E.J. Brouwer.
\newblock {\"U}ber die periodischen {T}ransformationen der {K}ugel.
\newblock {\em Math. Annalen}, 80:39--41, 1919.

\bibitem{Bredon72}
G.E. Bredon.
\newblock {\em Introduction to Compact Transformation Groups}.
\newblock Academic Press, New York, 1972.

\end{thebibliography}
\end{document}